\pdfoutput=1
\documentclass[a4paper,11pt]{article}

\usepackage{epsfig}
\usepackage{epstopdf}
\usepackage{amssymb}
\usepackage{amsthm}
\usepackage[cmex10]{amsmath}
\usepackage{amsmath}
\usepackage{amsfonts}
\usepackage{bbm}
\usepackage{multicol}
\usepackage{graphicx}
\usepackage[left=1.5cm,right=2.5cm]{geometry}
\usepackage{enumitem}
\usepackage{natbib}
\usepackage[colorlinks=true,linkcolor=blue, citecolor=blue, urlcolor=blue]{hyperref}
\usepackage[dvipsnames]{xcolor}
\usepackage{caption}
\usepackage[textfont=footnotesize,justification=centering]{subcaption}
\usepackage{multirow}
\usepackage{arydshln}
\usepackage{fancyhdr}
\usepackage{bm}
\usepackage{setspace}
\usepackage[hang,flushmargin]{footmisc}
\usepackage{mathtools}
\usepackage{bbm}
\usepackage{dsfont}
\usepackage{subcaption}
\usepackage{booktabs}
\usepackage{color}
\usepackage{hyperref}
\usepackage{colortbl}
\usepackage{setspace}
\usepackage{lipsum}
\usepackage{booktabs}

\newtheorem{theorem}{Theorem}[section]

\newtheorem{lemma}[theorem]{Lemma}

\newtheorem{assumption}[theorem]{Assumption}

\theoremstyle{definition}

\usepackage{chngcntr}
\counterwithin{equation}{section}

\def \bR{\mathbb{R}}
\def \bP{\mathbb{P}}
\def \bG{\mathbb{G}}

\def \cP{\mathcal{P}}
\def \cF{\mathcal{F}}
\def \cL{\mathcal{L}}
\def \cA{\mathcal{A}}
\def \cG{\mathcal{G}}
\def \cV{\mathcal{V}}

\def \cM{\mathcal{M}}

\def \E{\textrm{E}}
\def \VaR{\textrm{VaR}}

\def \vrmv{\textrm{vrmv}}
\def \srmv{\textrm{srmv}}

\def \static{\textrm{static}}

\def \exp{\textrm{exp}}
\def \pow{\textrm{pow}}
\def \mv{\textrm{mv}}
\def \sst{\textrm{st}}
\def \Rach{\textrm{Rach}}

\def \1{\mathbf{1}}

\def \mv{\textrm{mv}}

\usepackage{titling}
\usepackage{soul}
\usepackage{framed}
\soulregister\cite7
\soulregister\ref7
\soulregister\pageref7

\colorlet{shadecolor}{yellow}

\begin{document}
\begin{titlepage}

\title{\vspace*{-2cm} Mean-variance hybrid portfolio optimization with quantile-based risk measure  \thanks{This research is partially supported by the National Natural Science Foundation of China under the grants 71971132, 72201067, 71801088 and 72192832. This research is funded/supported by Shanghai Research Center for Data Science and Decision Technology.}
}

\date{}

\author{
Weiping Wu \thanks{School of Economics and Management, Fuzhou University, Fuzhou, Fujian, China. Email: wu.weiping@fzu.edu.cn}
\and Yu Lin \thanks{School of Economics and Management, Fuzhou University, Fuzhou, Fujian, China. Email: 210710020@fzu.edu.cn}
\and Jianjun Gao  \thanks{Corresponding author. School of Information Management and Engineering, Shanghai University of Finance and Economics, Shanghai, China. Email: gao.jianjun@shufe.edu.cn}
\and Ke Zhou \thanks{Business School, Hunan University, Changsha, Hunan , China. Email: kzhou@hnu.edu.cn}
}

\maketitle

\begin{abstract}
This paper addresses the importance of incorporating various risk measures in portfolio management and proposes a dynamic hybrid portfolio optimization model that combines the spectral risk measure and the Value-at-Risk in the mean-variance formulation. By utilizing the quantile optimization technique and martingale representation, we offer a solution framework for these issues and also develop a closed-form portfolio policy when all market parameters are deterministic. Our hybrid model outperforms the classical continuous-time mean-variance portfolio policy by allocating a higher position of the risky asset in favorable market states and a less risky asset in unfavorable market states. This desirable property leads to promising numerical experiment results, including improved Sortino ratio and reduced downside risk compared to the benchmark models.
\end{abstract}



\end{titlepage}

\newpage
\section{Introduction} \label{sec_introduction}

The mean-variance (MV) portfolio selection model, introduced by \cite{Markowitz:1952}, is widely used in academic research and investment practice. However, the symmetric nature of the variance term penalizes both gains and losses, making it a drawback of the MV formulation. To deal with this issue, various risk measures have been introduced in the portfolio optimization model. The concept of \textit{coherent risk measure} (\citealp{ArtznerDelbaenEberHeath:1999}) and its extension, the convex risk measure (\citealp{Follmer:convexrisk2002}) set up the basic principles for defining reasonable risk measures. Along with the development of modern risk measures, the mean-risk portfolio decision models have been extensively studied under the frameworks of static and dynamic portfolio optimizations (e.g., see \citealp{Kolm:2014, GaoZhouLiCao:2017,ZhouGaoLiCui:2017, HeZhou:2015, Ortobelli:2008, AdamHoukariLaurent:2008, Steuer:EJOR,Roman:2009}). Among these researches, one particular direction focuses on the portfolio optimization model with multiple risk measures (e.g., see \citealp{GaoXiongLi:2016, Steuer:EJOR, RomanDarbyMitra:2007, Roman:2009}). This paper follows similar spirits to study the dynamic mean-risk portfolio decision model with both MV formulation and the quantile-based risk measure.

We combine two types of quantile-based risk measures in the continuous-time MV formulation, namely, the \textit{Spectral Risk Measure} (SRM) and the \textit{Value-at-Risk} (VaR). SRM, proposed by \cite{Acerbi:2002}, calculates risk as a weighted average of the quantiles of the return distribution (wealth). When the weighting function (also called the Spectrum) satisfies certain mild conditions, SRM becomes a coherent risk measure (e.g., see \citealp{AdamHoukariLaurent:2008,Acerbi:2002}). Despite being a special case of the distortion risk measure \cite{AdamHoukariLaurent:2008}, SRM's convexity and flexibility in choosing the spectrum function make it a useful choice for constructing the portfolio optimization model.  Besides SRM, this work also considers the dynamic MV formulation together with the VaR as the additional risk measure. VaR is traditionally considered a standard downside risk measure since it measures the quantile of the loss distribution. However, VaR has some drawbacks in portfolio optimization, such as its inability to diversify risk and nonconvexity of the problem. To overcome these limitations, \cite{RockafellarUryasev:2000,RockafellarUryasev:2002} propose a Conditional Value-at-Risk (CVaR)-based portfolio decision model and develop an LP-based solution scheme. However, recent studies such as \cite{LimShanthikumarVahn:2011} indicate that the CVaR-based portfolio optimization model is highly sensitive to estimation errors. On the other hand, VaR is shown to be more robust than convex risk measures like CVaR by \cite{ContDeguestScandolo:2010,Cont:2013,KouPengHeyde:2013}. Additionally, \cite{KouPengHeyde:2013} demonstrate that VaR is a more appropriate risk measure for imposing trading book capital requirements.

Based on the continuous-time MV formulation, we propose two novel portfolio optimization models with multiple risk measures, namely, the dynamic mean-variance-Spectral Risk Measure (SRM-MV) and mean-variance-Value-at-Risk (VaR-MV) models. Incorporating these risk measures in the MV formulation is highly meaningful as it allows for better shaping of the probability distribution of the terminal wealth. In particular, the introduction of these risk measures strengthens the management of downside risk in the loss domain. A recent study by \cite{Staden:SIAMFM2021} indicates that the distribution of terminal wealth generated by the dynamic MV policy may have a long tail in the loss domain. This phenomenon is further illustrated in our example presented in Figure \ref{fig:pdf_srmv_exp}. As a result, it becomes crucial to incorporate the downside risk in the MV formulation. On the other side, these downside risk measures reduce the relative importance of the variance in the objective function, which helps to mitigate the conservatism caused by the variance in the domain of gain. Another advantage of incorporating MV formulation with downside risk is that it can overcome some ill-posedness issues. Numerous research including \cite{JinYanZhou:2005,HeZhou:2015,GaoZhouLiCao:2017}) have pointed out that a large class of continuous-time mean-downside risk portfolio optimization (including CVaR,VaR, Weighted-VaR) is ill-posed in sense that the wealth becomes unbounded. However, when the variance is included in such a model, the variance term will naturally prevent the wealth to go to infinity.

This paper not only presents innovations in portfolio optimization modeling through the introduction of two dynamic portfolio optimization models with multiple risk measures but also makes several other significant contributions. Firstly, we propose a solution scheme that combines the martingale representation with the quantile-based optimization technique to solve these problems. This approach broadens the scope of these methods, which were originally developed for the behavioral portfolio model (e.g., see \citealp{JinZhou:2008,He2011,HeZhou:2011}). Furthermore, we derive closed-form solutions for these models in a Black-Scholes-type market setting. The explicit expression of such a portfolio policy enables us to examine the key difference brought by the downside risk measure in comparison to the MV portfolio policy. For a more general market setting with stochastic returns rate or volatility, we propose a partial differential equation-based approach. Thirdly, we conduct numerical tests to demonstrate the effectiveness of our proposed portfolio models in controlling downside risk and improving the Sortino Ratio (\citealp{Sortino:2001}). As a byproduct, we also develop the solution scheme for the static SRM-MV and VaR-MV portfolio optimization problem, which is provided in  \ref{appendix_static}. These models serve as benchmarks in the numerical test.

\subsection{Related Literature and paper structure}\label{sse:literature}

This research is related to the search for a mean-multiple-risks portfolio optimization model. In the context of static portfolio optimization, \cite{RomanDarbyMitra:2007} study a model that combined variance and CVaR as risk measures. By using the parametric representation of CVaR (see Rockafellar and Uryasev, 2000), this problem can be reformulated as a convex QP. \cite{Cesarone:mv-var-2021} investigate VaR-MV portfolio optimization and demonstrated that its out-of-sample performance is better than the equally weighted portfolio and MV-CVaR portfolio. \cite{Utz:EJOR2014} study the inverse fund optimization problem with a multi-objective function. However, there are few reports in the literature on the dynamic version of this type of portfolio model. In the continuous-time market setting, Gao et al. (2016) examined portfolio optimization models that incorporated both MV formulation with CVaR or the safety-first principle. Using the martingale approach, this work provided a solution scheme for these problems. In addition to the mean-risk formulation, some research has studied the continuous-time utility maximization problem with additional risk constraints, such as the VaR, Safety-First-Principle, or the variance (see, e.g., \citealp{BasakShapiro:2001,Bensoussan:2022,ChiuWongZhao:2018}).

The static SRM-based portfolio optimization model has been studied by \cite{Acerbi:2002}, who showed that such a problem could be formulated as a linear programming problem when uncertainty was represented by discrete scenarios. \cite{AdamHoukariLaurent:2008} evaluate the performance of the SRM-based model, while \cite{AbadIyengar:2015} consider a portfolio model with multiple SRMs. Recently, \cite{GuoXu:2022} extended these models to the robust SRM formulation and developed a tractable solution scheme. In terms of SRM-based continuous-time portfolio optimization, our work is the first to study such a problem. Current research is also related to VaR-based portfolio optimization. \cite{ZhouGaoLiCui:2017} is the first to investigate mean-VaR portfolio optimization in a continuous-time setting. As the problem has an ill-posed issue, they introduced an artificial upper bound for the wealth. A similar approach was adopted to solve the mean-Safety-First portfolio optimization (\citealp{ChiuWongLi:2012}) and the Weighted-VaR-based model (\citealp{HeZhou:2015}). Additionally, another strand of research attempts to extend the static SRM to continuous-time dynamic spectral risk measure (\citealp{MadanPistoriusStadje:2017}).

Our research also advances the research on the continuous-time MV (CTMV) portfolio selection. Since the publication of seminar works by \cite{ZhouLi:2000}, the CTMV portfolio optimization model has been extensively explored, with notable studies including \cite{BieleckiJinPliskaZhou:2005,Chiu:EJOR2012,vanDangForsyth:2021,Staden:SIAMFM2021}. Given that variance may lead to time-inconsistent issues, a significant portion of research has focused on developing time-consistent policies for CTMV portfolio optimization (e.g., see \citealp{Wang:EJOR-TCMV-2011,basak2010dynamic,dang2016better}). Notably, the dynamic SRM (or VaR)-based portfolio optimization model also has time-consistency issue, and hence the policy derived in this work falls under the category of pre-committed policy. Developing time-consistent policies for SRM-MV and VaR-MV portfolio optimization is beyond the scope of our current paper.

The remainder of this paper is structured as follows. In Section \ref{sec:model}, we introduce the market model and the hybrid portfolio decision models. The solutions for these hybrid portfolio models are developed in Section \ref{se:sol_portfolio}. In Section \ref{sec_example}, we study the properties of the hybrid portfolio policies and evaluate the performance of different models. We conclude the paper in Section \ref{sec_conclusion}. Throughout the paper, we use $\1_{{\mathcal{A}}} $ to denote the indicator function, which equals $1$ if the condition $\mathcal{A}$ holds and $0$ otherwise. The notation $B^{\top}$ represents the transpose of matrix (or vector) $B$. The probability density function and the cumulative distribution function (CDF) of a standard normal variable are denoted as $\phi(\cdot)$ and $\Phi(\cdot)$, respectively. The solution scheme for static SRM-based model is provided in in  \ref{appendix_static}.

\section{Market model and problem formulations} \label{sec:model}

\subsection{Quantile-based risk measure}\label{sse:risk_measure}
A broad class of quantile-based risk measures can be represented by the integration of some weighted quantile functions defined for the random loss (e.g., see \citealp{Acerbi:2002,AdamHoukariLaurent:2008,DowdCotterSorwar:2008}). This formulation of risk measures encompasses popular measures such as Value-at-Risk (VaR) and Expected Shortfall (ES or CVaR) as special cases. For ease of illustration, we define the quantile-based risk measure as follows. Suppose the investment horizon is $T$, and the terminal wealth of the investment is denoted by $x(T)$. Given a probability space with probability measure $\mathbb{P}(\cdot)$, we use $F(y) \triangleq \bP\big( x(T)\leq y \big)$ and $G(s) \triangleq \inf \{y\in\bR~|~F(y)>s\}$ to denote the cumulative distribution function and the upper quantile function of $x(T)$, respectively. Following a similar definition in \cite{Acerbi:2002}\footnote{Note that, in this work, we consider the loss of investment as $-x(T)$.}, the quantile-based risk measure is defined as follows,
\begin{align}
\mathcal{M}_{\psi}[x(T)] \triangleq -\int_{0}^{1}  \psi(s) \cdot G(s)ds, \label{def_quantile_risk}
\end{align}
where $\psi(\cdot):[0,1] \rightarrow \mathbb{R}_+$ is a user-defined weighting function satisfying $\int_0^1 \psi(s)ds=1$. By choosing different weighting functions $\psi(\cdot)$, the quantile-based formulation in Eq. (\ref{def_quantile_risk}) can represent various commonly used risk measures. In this work, we focus on two types of risk measures: the \textit{Spectral Risk Measure} and the \textit{Value-at-Risk}.

\begin{figure}[h]
\centering
\includegraphics[width=180pt]{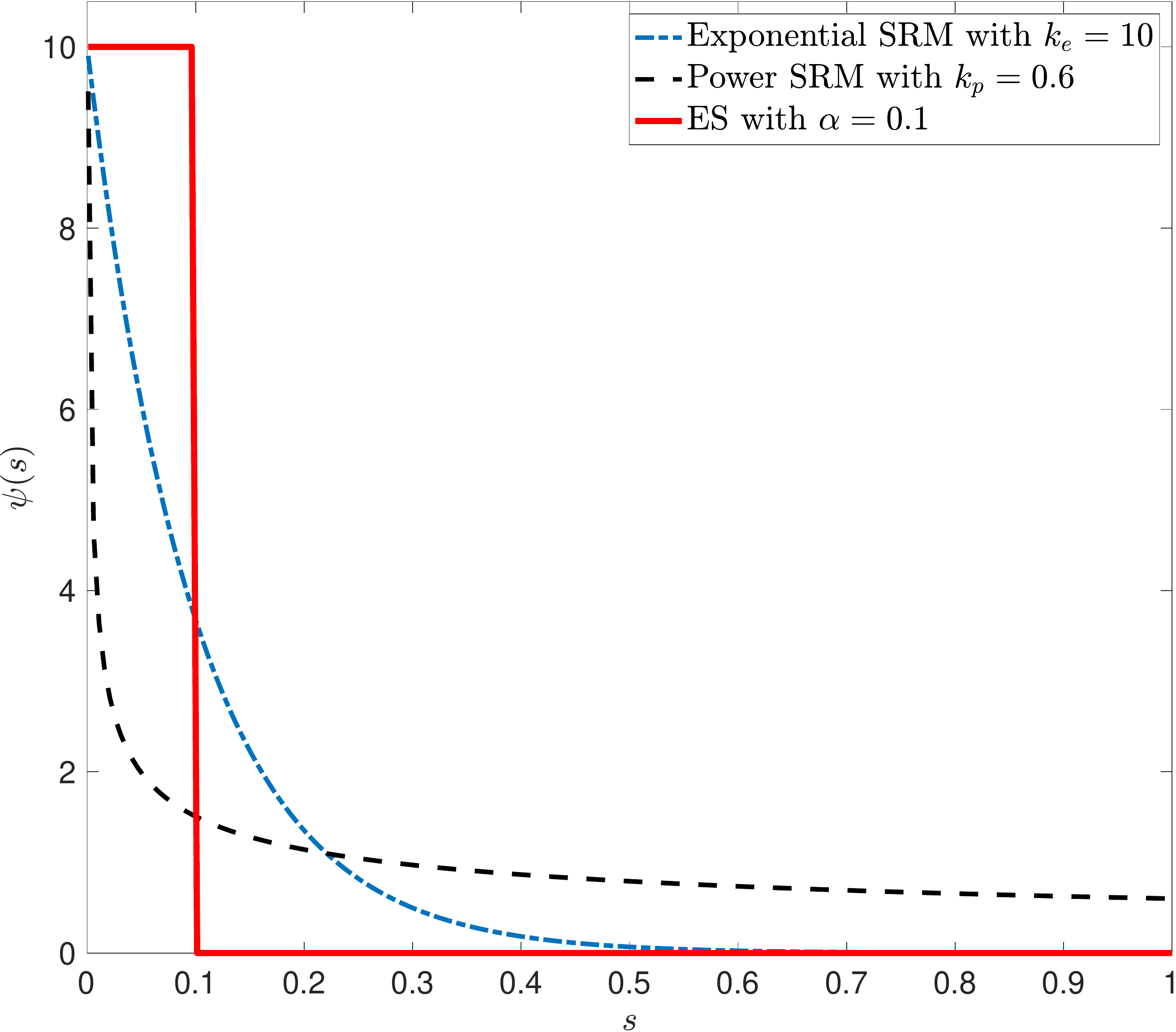}
\caption{The spectrum function $\psi(s)$ in Spectral Risk Measure}
\label{fig_psi}
\end{figure}

\textbf{Spectral Risk Measure:} In formulation (\ref{def_quantile_risk}), if the weighting functions $\psi(\cdot)$ is non-negative, non-increasing and right-continuous, then risk measure $\mathcal{M}_{\psi}[x(T)]$ becomes the \textit{Spectral Risk Measure} (SRM). The SRM plays an important role in the modern risk measures theory, as it is {a coherent}, comonotonic additive and low-invariant risk measure (see \citealp{Kusuoka:2001,Acerbi:2002,Brandtner:2016}). In SRM, the weighting function $\psi(\cdot)$ is commonly referred to as the spectrum. If we set the spectrum as a step function, i.e., $\psi(s)=\frac{1}{\gamma} \1_{{ 0\leq s\leq \gamma } }$ in (\ref{def_quantile_risk}), the resulting risk measure becomes $\cM_{\psi}[x(T)] = - \frac{1}{\gamma}\int_{0}^{\gamma} G(s) ds$, which is known as the $\gamma$-level \textit{Expected Shortfall} (ES). Besides the ES, there are other ways to specify the spectrum. Based on the utility theory \footnote{\cite{Bertsimas:JEDC2004} reveal that we may use the utility function to design the suitable spectrum functions in SRM.}, it is possible to define the exponential and power functions-based spectrum (\citealp{DowdCotterSorwar:2008}) as $\psi(s)\triangleq\frac{k_e \cdot e^{-k_e\cdot s}}{1-e^{ -k_e }}$ and
$ \psi(s)\triangleq k_p\cdot s^{k_p-1}$ where $k_e \in (0,\infty)$ and $k_p\in(0,1]$ are the parameters (see, e.g., \citealp{Brandtner:2013,DowdCotterSorwar:2008}). Figure \ref{fig_psi} displays the typical shapes of the exponential, power, and step-function-based spectra. One crucial characteristic of these spectral functions is that they assign more weight to smaller probabilities than larger ones.

\textbf{Value-at-Risk:} Given a confidence level $\gamma \in (0,1)$, the $\gamma$-VaR of the terminal wealth $x(T)$ is usually defined as (\citealp{FollmerSchied:2004}), $\VaR_{\gamma}[x(T)] $$\triangleq$$ \inf \{~y \in \bR ~|~ \bP(- x(T) \leq y) \geq 1-\gamma\}$ which measures the maximal value of the `loss' (i.e., $-x(T)$) with a given probability $1-\gamma$. \footnote{Note that the confidence level $\gamma$ is usually set at small values as $\gamma=1\%$, $5\%$, $10\%$.} VaR is also a special case of the formulation (\ref{def_quantile_risk}), if we set the weighting function as the \textit{Dirac delta function}\footnote{Function $\delta(x)$ is called the Dirac delta function, if it satisfies: $\delta(x)=0$ when $x\not=0$; $\delta(x)=+\infty$ when $x=0$, and $\int_{-\infty}^{+\infty}\delta(s) ds=1$. One important property is that, given $f(x):\mathbb{R}:\rightarrow \mathbb{R}$, it has $\int_{-\infty}^{\infty}f(s)\delta(t-s) ds=f(t)$. } as $\psi(s) = \delta(s-\gamma)$, then it has
\begin{align}
\VaR_{\gamma}[x(T)]=\cM_{\psi}[x(T)]= -G(\gamma), \label{def_var_quantile}
\end{align}
where the last equality is from the property of the Dirac delta function.

Although VaR and SRM share a similar quantile-based formulation (\ref{def_quantile_risk}), they are fundamentally distinct. The primary difference lies in the requirement of the spectrum function $\psi(\cdot)$ in SRM to be non-increasing and right-continuous. As VaR does not possess these properties, it fails to satisfy sub-additivity, and hence, it is not a coherent risk measure. Consequently, the resulting portfolio optimization problem is not a convex optimization problem. Due to this crucial difference, we need to treat SRM-based and VaR-based portfolio optimization problems separately using different methods in the subsequent sections.

\subsection{Continuous-time market and dynamic portfolio optimization models}\label{sse:continuous-model}
We consider a financial market with one risk-free asset and $n$ risky assets, which are traded continuously within a finite horizon $[0,T]$. All randomness of the market is modeled by a complete filtrated probability space $(\Omega,\cF,\bP,\{\cF_t\}_{t\geq0})$. On this space, we introduce $n$-dimensional Brownian motion $W(t)$ $=$ $(W^1(t),\cdots,W^n(t))^{\top}$ and assume that $W^i(t)$ and $W^j(t)$ are mutually independent for all $i\neq j$. We use $\cF_t$ to denote the information set available at time $t\in[0,T]$.\footnote{ Formally, the $\cF_t$ is called the filtration, which is the augmented $\sigma$-algebra generated by the paths of $W(t)$.} Let $S_0(\cdot)$ be the price process of the risk-free asset which satisfies the differential equation, $dS_0(t) = r(t) S_0(t) dt$ with $S_0(0)=s_0>0$, where $r(\cdot)$ is the risk-free return rate. Let $S_i(\cdot)$ be the price process of the $i$-th risky asset, which is governed by the following stochastic differential equation,
\begin{align}
dS_i(t) ~=~ S_i(t)\big(\mu_i(t) dt +\sum_{j=1}^{n}\sigma_{ij}(t) dW^j(t)\big), \label{def_St}
\end{align}
with $S_i(0)=s_i>0$ for $t\in[0,T]$ where $\mu_i(\cdot)\in \mathbb{R}$ and $\sigma_{ij}(\cdot)\in \mathbb{R}$ are the appreciation rate and volatility, respectively, for $i=1,\cdots,n$. We assume that the volatility matrix $\sigma(t)\triangleq\{\sigma_{ij}(t)\}_{i,j=1}^{n}\}$ satisfies the nondegeneracy condition, i.e., $\sigma(t)\sigma(t)^{\top}$ is positive definite almost surely for $t\in[0,T]$. Furthermore, we assume that $r(t)$, $\{\mu_i(t)\}_{i=1}^n$ and $\{\sigma_{ij}(t)\}_{i,j=1}^{n,n}$ are scalar-valued $\cF_t$-measurable and uniformly bounded stochastic processes for any $t\in[0,T]$.

An investor enters the market with initial wealth $x_0>0$ and allocates his/her wealth continuously on these assets. For $t\in[0,T]$, let $x(t)$ be the total wealth level at time $t$ and $u(t) \triangleq \big(u_1(t),\cdots,u_n(t)\big)^{\top}$ be the portfolio allocation at time $t$ where $u_i(t)$ represents the wealth allocated on $i$-th risky asset at time $t$. Under the self-financing portfolio policy, the wealth dynamics satisfy the following dynamics,
\begin{align}
dx(t) = \big( r(t)x(t)+b(t)^{\top}u(t) \big)dt
   + u(t)^{\top}\sigma(t)dW(t),~~t\in[0,T],  \label{def_xt}
\end{align}
where $x(0)=x_0$ and $b(t)\triangleq \mu(t)-r(t)=(\mu_1(t)-r(t),\cdots,\mu_n(t)-r(t))^{\top}$ denotes the excess return rate for $t\in[0,T]$.

We use $\E[x(T)]$ and $\cV[x(T)] \triangleq \E[(x(T)-E[x(T)])^2]$ to denote the unconditional expected value and the variance of the terminal wealth $x(T)$. To control both spectral risk and the variance of the terminal wealth, the investor considers the following SRM-MV hybrid portfolio optimization model,
\begin{align}
(\cP_{\srmv}):~&~\min_{u(t), t\in[0,T]}~\cV[x(T)] + \omega_{\srmv} \cdot \mathcal{M}_{\psi}[x(T)] \notag\\
(s.t.)~&~\E[x(T)]=x_d, \notag \\
      ~&~(x(\cdot), {u(\cdot)})~\textrm{satisfies}~(\ref{def_xt}), \notag \\
      ~&~x(T)\geq0, \label{const_no_bankrupt}
\end{align}
where {$x_d>0$ is the expected terminal wealth level}, $\mathcal{M}_{\psi}[\cdot]$ is the spectral risk measure and $\omega_{\srmv}\geq0$ is the weighting parameter controlling the importance of {SRM}. To ensure tractability, we assume the risk spectrum function $\psi(\cdot)$ satisfies the following condition.
\begin{assumption}\label{asmp:SMR_psi}
The spectrum function $\psi(\cdot):[0,1]:\rightarrow \mathbb{R}$ satisfies: (i) $\psi(s)\geq 0$, (ii) $\psi(s)$ is differentiable, (iii) $d \psi(s)/ds \leq 0$ for all $s\in[0,1]$, and (iv) $\int_{0}^{1} \psi(s)ds=1$.
\end{assumption}
Besides the model $(\cP_{\srmv})$, we are also interested in integrating the VaR in the dynamic MV portfolio optimization model, i.e., we consider the following VaR-MV hybrid portfolio optimization model,
\begin{align*}
(\cP_{\vrmv}):~~&~\min_{u(t), t\in[0,T]}~\cV[x(T)] + \omega_{\vrmv} \cdot \VaR_{\gamma}[x(T)] \\
(s.t.)~&~\E[x(T)]=x_d, \notag \\
      ~&~(x(\cdot),{u(\cdot)})~\textrm{satisfies}~(\ref{def_xt}), \\
      ~&~x(T)\geq0,
\end{align*}
where $\omega_{\vrmv}\geq0$ is the weighting parameter that balances the importance of the variance and VaR. Problem $(\cP_{\srmv})$ and $(\cP_{\vrmv})$ are an extension of the conventional dynamic MV portfolio decision model (e.g., see \cite{BieleckiJinPliskaZhou:2005}, \cite{ZhouLi:2000}) where additional risk measures are included. The VaR-MV portfolio optimization problem  $(\cP_{\vrmv})$ is also an extension of the model studied in \cite{ZhouGaoLiCui:2017} in which only the VaR is included as the risk measure.\footnote{It is worthwhile to mention that solely using the VaR as a risk measure in a continuous-time portfolio model may have an ill-posedness issue (see, e.g., \cite{JinYanZhou:2005}, \cite{HeZhou:2015}, \cite{ZhouGaoLiCui:2017}). To deal with this issue, \cite{ZhouGaoLiCui:2017} and {\cite{GaoZhouLiCao:2017}} propose to add the artificial upper bound on the terminal wealth. However, in formulation, $(\cP_{\vrmv})$, this artificial bound is no longer needed since the variance term tames the terminal wealth to take the finite value.}

In the consequent sections, if it is necessary, we add a subscript to distinguish solutions for different models, i.e.,
$\{x^*_{\srmv}(t),u^*_{\srmv}(t)\}|_{t=0}^T$ and $\{x^*_{\vrmv}(t),u^*_{\vrmv}(t)\}|_{t=0}^T$ denote the optimal wealth and portfolio for model $(\cP_{\srmv})$ and  $(\cP_{\vrmv})$, respectively.

\section{Optimal solution of hybrid portfolio model}\label{se:sol_portfolio}

In this section, we develop the solutions for models ($\cP_{\srmv}$) and ($\cP_{\vrmv}$). Although these problems can be viewed as stochastic optimal control problems (\citealp{YongZhou:1999, Pham:2009}), due to the complicated constraints and quantile-based objective function, it is hard to apply the classical optimal control approach directly. Instead, we take advantage of the completeness of the market model and adopt the martingale method (e.g., see \citealp{KaratzasShreve:1998, JinZhou:2008, BieleckiJinPliskaZhou:2005}) to solve these problems. Specifically, we take two steps to solve these problems: (i) characterizing the optimal terminal wealth $x^*(T)$ by identifying its optimal quantile function (e.g. see \cite{JinZhou:2008,HeZhou:2011,HeZhou:2015}); (ii) and developing the optimal portfolio policy to replicate such an optimal terminal wealth.

\subsection{Optimal terminal wealth} \label{sec_terminal_wealth_SRM}
Due to the complete market model, we may define the state price density process (SPD) $z(t)$ as $ z(t) = -z(t)\big( r(t) dt + \theta(t)^{\top}dW(t) \big)$ with $z(0)=1$, where $\theta(t)\triangleq \sigma(t)^{-1}b(t)$ is risk premium process for $t\in[0,T]$ (see \cite{KaratzasShreve:1998}).\footnote{The stochastic discount factor $z(t)$ is also called the state-price deflator process in \cite{Duffie-book}.} Equivalently, $z(t)$ can be expressed as,
\begin{align}
z(t)=\exp\Big\{ -\int_0^t (r(\nu)&+\frac{1}{2}\|\theta(\nu)\|^2)d\nu- \int_0^t\theta(\nu)^{\top}dW(\nu) \Big\}.\label{def_zt}
\end{align}
Using the SPD, the discounted wealth process $z(t)x(t)$ becomes a martingale (\cite{KaratzasShreve:1998,Duffie-book}), i.e., it has
\begin{align}
z(t)x(t)=\E\big[z(T)x(T)~|~\cF_t~\big], \label{def_xt_expectation}
\end{align}
for any $0\leq t \leq \tau\leq T$. Without loss of generality, we assume the following condition is true.\footnote{Imposing this condition guarantees the monotonicity property of the quantile function. A similar condition is also imposed \cite{JinYanZhou:2005}.}
\begin{assumption}\label{asmp_zt}
The random variable $z(T)$ admits no atom, i.e., $\bP(z(T)=a)=0$ for any $a\in(0,\infty]$.
\end{assumption}

To solve the terminal wealth of problem $(\cP_{\srmv})$, we define the distribution function  of $z(T)$ as $K_0(y)\triangleq \E[\1_{ \{z(T)\leq y\} }] = \mathbb{P}(z(T)\leq y)$ for some $y \in \mathbb{R}$. Since $z(T)\geq 0$, it has $K_0(y)=0$ when $y<0$. Under the Assumption \ref{asmp_zt}, we may define the inverse function of $K_0(\cdot)$ as $K_0^{-1}(s)$$\triangleq$$\inf\{ z \in \bR~|~K_0(z)>s \}~\textrm{for}~s\in[0,1]$. Using the embedding method (e.g., see \citealp{LiNg:2000,ZhouLi:2000,BieleckiJinPliskaZhou:2005}), we may rewrite the variance term in problem ($\cP_{\srmv}$) as $\cV[x(T)]= \E[(x(T)-x_d)^2]$ where $x_d=\E[x(T)]$ is a parameter. Such a formulation provides a quadratic objective function in the MV-based portfolio optimization. Then, using the martingale property, we may solve the following auxiliary problem for the optimal terminal wealth $x^*(T)$ of the problem ($\cP_{\srmv}$), \footnote{Notation $\cL_{\cF_T}^2(\Omega;\bR)$ means the set of all $\bR$-valued $\cF_T$-measurable random variables.}
\begin{align}
(\cA_{\srmv}):\min_{ x(T)\in \cL_{\cF_T}^2(\Omega;\bR)}~~&~\E[x^2(T) - x_d^2] + \omega_{\srmv}\cdot\cM_{\psi}[x(T)] \notag\\
(s.t.)~~&~\E[x(T)] = x_d, \label{const_beta}\\
      ~~&~\E[z(T)x(T)] = x_0, \label{const_x0}\\
	  ~~&~x(T)\geq0, \label{const_nobank}
\end{align}
where the constraint (\ref{const_x0}) is from (\ref{def_xt_expectation}).

As the objective function of problem $(\cA_{\srmv})$ involves SRM, it is more convenient to adopt the quantile formulation (e.g., see \citealp{JinZhou:2008,HeZhou:2011,HeZhou:2015}) to solve such a problem. We will utilize the notation introduced earlier, where $F(\cdot)$ and $G(\cdot)$ represent the distribution function and quantile function, respectively, of the random terminal wealth $x(T)$. In the quantile approach, the main idea is to replace the decision variable $x(T)$ by its quantile function $G(\cdot)$. Specifically, the expected value and the second-order moment of $x(T)$ can be expressed as $\E[x(T)]$$=$$\int_{-\infty}^{\infty} u dF(u)$$=$$\int_{0}^{1} G(s) ds$ and $\E[x(T)^2]$ $=$ $\int_{-\infty}^{\infty} u^2 d F(u)=\int_{0}^1 G^2(s) ds$. The constraint (\ref{const_nobank}) is equivalent to $G(s)\geq 0$ for $0\leq s \leq 1$. As for constraint (\ref{const_x0}), we may employ  Theorem B.1 in \cite{JinZhou:2008} (under Assumption \ref{asmp_zt}), which can be expressed as $\E[x(T)z(T)] =\E[G(1-K_0(z(T)))z(T)]$. Note that, $z(T)$ can be expressed as $z(T)=K_0^{-1}(1-s)$ with $s=1-K_0(z(T))$, we further obtain $\E[x(T)z(T)]=\E[G(s)K_0^{-1}(1-s)]$. Since $K_0(\cdot)$ is the CDF of $z(T)$, we know that $s=1-K_0(z(T))$ follows the uniform distribution in $[0,1]$, which further yields,
\begin{align*}
\E[x(T)z(T)]=\E[ G(s)K_0^{-1}(1-s)]=\int_{0}^{1} G(s) K_{0}^{-1}(1-s) ds.
\end{align*}
Then the quantile function-based formulation of problem $(\cA_{\srmv})$ is as follows,
\begin{align}
(\cG_{\srmv}):~&~\ \min_{ G(\cdot) \in \bG}~ \int_{0}^{1} G^{2}(s) ds - \omega_{\srmv} \int_{0}^{1} \psi(s)G(s) ds \notag \\
(s.t.)~&~ \int_{0}^{1} G(s) K_{0}^{-1}(1-s) ds = x_0, \label{const_Gt_x0}\\
      ~&~ \int_{0}^{1} G(s) ds = x_d, \label{const_Gt_d} \\
      ~&~ G(s)\geq0,~~0\leq s\leq1, \label{const_Gt_bankruptcy}
\end{align}
where the set $\bG$ denotes the feasible set of the quantile functions,
\begin{align*}
\bG \triangleq\{G(\cdot):& [0,1]\rightarrow[0,\infty] ~|~G(\cdot)~\textrm{is nondecreasing} ~\textrm{and right-continuous function} \}.
\end{align*}
Problem $(\cG_{\srmv})$ is a convex functional optimization problem that can be solved analytically. The following result characterizes the optimal solution of problem $(\cG_{\srmv})$ and problem ($\cA_{\srmv}$).

\begin{theorem} \label{thm_srmv_xT}
The optimal solution of problem ($\cG_{\srmv}$) is $G_{\srmv}^*(s)=\frac{1}{2}\Big(\rho^* - \eta^*   K_0^{-1}(1-s)+\omega_{\srmv}\psi(s)\Big) \1_{\{s^{\dag}\leq s\leq 1\}}$,
and the optimal terminal wealth of problem ($\cA_{\srmv}$) is
\begin{align}
x_{\srmv}^*(T)= \frac{1}{2} \Big(  \rho^* - \eta^* z(T) + \omega_{\srmv} \cdot \psi(1 - K_0(z(T))) \Big) \1_{ \{ 0 < z(T) \leq z^{\dag} \} } ,
\label{def_srmv_xT}
\end{align}
where $s^{\dag}$ is the solution of the following equation,
\begin{align}
\rho^* - \eta^* K_0^{-1}(1-s^{\dag}) + \omega_{\srmv}\psi(s^{\dag}) = 0,\label{equ_srmv_s_dag}
\end{align}
and $\rho^*$ and $\eta^*>0$ are the solution of the following system of two equations,
\begin{align}
&\rho^*-\eta^* z^{\dag} + \omega_{\srmv}\psi\big(1- K_0(z^{\dag})\big) = 0, \label{equ_zdag_sr}\\
&2x_0 = \E \Big[ \big( \rho^*-\eta^* z(T) + \omega_{\srmv}\psi(1 - K_0(z(T))) \big) z(T) \1_{  \left\{ 0 < z(T) \leq z^{\dag} \right\} } \Big], \label{equ_x0_sr} \\
&2x_d=  \E \Big[ \big( \rho^*-\eta^* z(T) + \omega_{\srmv}\psi(1 - K_0(z(T))) \big)\1_{  \left\{ 0 < z(T) \leq z^{\dag} \right\} } \Big]. \label{equ_xd_sr}
\end{align}
\end{theorem}
In Theorem \ref{thm_srmv_xT}, the optimal solution of problem ($\cG_{\srmv}$) is represented by the quantile function $G_{\srmv}^*(\cdot)$ which is further translated to the optimal wealth as given in Eq. (\ref{def_srmv_xT}) through the relationship $ x^*(T) = G^*(1-K_0(z(T)))$ (see, e.g., Theorem B1 in \cite{JinZhou:2008}).

The formulation (\ref{def_srmv_xT}) indicates that the optimal wealth generated from the classical dynamic MV portfolio selection model is just a special case (see Theorem 4.1 in \cite{BieleckiJinPliskaZhou:2005}) of (\ref{def_srmv_xT}). Indeed, if we set $\omega_{\srmv}=0$ in (\ref{equ_zdag_sr}), (\ref{equ_x0_sr}) and (\ref{equ_xd_sr}), then it yields the optimal terminal wealth of dynamic MV model as
\begin{align}
x_{\mv}^*(T) = \frac{1}{2}\big( \hat{\rho}^*-\hat{\eta}^* z(T)\big)\1_{ \{ 0 < z(T) \leq \hat{\rho}^*/\hat{\eta}^*\} }, \label{def_mv_xT}
\end{align}
where $\hat{\rho}^*$ and $\hat{\eta}^*$ are the parameters for the case $\omega_{\srmv}=0$. Comparing $x^*_{\mv}(T)$ and $x^*_{\srmv}(T)$, we have the following decomposition,
\begin{align}
x_{\srmv}^*(T)=
\underbrace{ \frac{\rho^*-\eta^* z(T)}{2} \1_{ \{0 < z(T) \leq z^{\dag} \} }}_{ \textrm{MV} }
+\omega_{\srmv}\underbrace{\frac{\psi\big( 1 - K_0(z(T)) \big)}{2}\1_{\{0 < z(T) \leq z^{\dag}\}} }_{ \textrm{SRM} }.\label{def_srmv_xT_decompose}
\end{align}
Such a decomposition means that the optimal terminal wealth of the SRM-MV model is the weighted summation of the MV model's optimal wealth and the spectrum function nested by the distribution function of $z(T)$.

We then turn to the VaR-MV hybrid portfolio model $(\cP_{\vrmv})$. Before we go forward, we want to point out that $\VaR_{\gamma}[x(T)]$ has some lower bound, $\underline{\beta} \leq \VaR_{\gamma}[x(T)] \leq 0$, where $\underline{\beta} \triangleq -\frac{x_0}{(K_1(K_0^{-1}(1-\gamma))}$ {with $K_1(y) \triangleq \E[z(T) \1_{ \{z(T)\leq y\} }]$}. In the above formulation, the lower bound is given in Proposition 3.2 in \cite{ZhouGaoLiCui:2017} and the upper bound is from the fact $x(T)\geq 0$ in problem $(\cP_{\vrmv})$. These bounds mean that, no matter what portfolio policy {$u(\cdot)$} is chosen, the VaR value $\VaR_{\gamma}[x(T)]$ generated in problem $(\cP_{\vrmv})$ is always in the interval $[\underline{\beta}, 0]$.

Using the martingale property, we may characterize the optimal terminal wealth for problem  $(\cP_{\vrmv})$ from the following auxiliary problem $(\cA_{\vrmv})$,
\begin{align*}
(\cA_{\vrmv}):\min_{x(T)\in \cL_{\cF_T}^2(\Omega;\bR)}~~&~\E[x^2(T)-x_d^2] + \omega_{\vrmv} \cdot \VaR_{\gamma}[x(T)] \\
(s.t.)~~&~x(T)~\textrm{satisfies}~(\ref{const_beta}),(\ref{const_x0})~\textrm{and}~(\ref{const_nobank})
\end{align*}
Similar to problem ($\mathcal{G}_{\srmv}$), it is more convenient to reformulate the problem $(\cA_{\vrmv})$ in quantile formulation as follows,
\begin{align}
(\cG_{\vrmv}):~&~\ \min_{ G(\cdot) \in \bG}~ \int_{0}^{1} G^{2}(s) ds - \omega_{\vrmv}\cdot G(\gamma) \label{G_mvv_objective} \\
(s.t.)~&~ G(s)~\textrm{satisfies}~(\ref{const_Gt_x0}),(\ref{const_Gt_d})~\textrm{and}~(\ref{const_Gt_bankruptcy}) \notag
\end{align}
where the term $\VaR_{\gamma}[x(T)]$ is represented by (\ref{def_var_quantile}). Solving problem ($\mathcal{G}_{\vrmv}$) gives the optimal quantile function\footnote{The optimal quantile function of problem {($\mathcal{G}_{\vrmv}$)} is given in { \ref{appendix_proof_vrmv_xT}} } which can be further translated to the corresponding optimal terminal wealth for the problem ($\cA_{\vrmv}$).

\begin{theorem} \label{thm_vrmv_xT}
The optimal solution of problem ($\cA_{\vrmv}$) is
\begin{align}
x_{\vrmv}^*(T)= \frac{\rho^* - \eta^* z(T)}{2} \1_{ \{ K_0^{-1}(1-\gamma) <z(T)\leq C_1 \}}
-\beta^* \1_{\{ C_2<z(T)\leq K_0^{-1}(1-\gamma)\} }
+ \frac{\rho^*-\eta^* z(T)}{2} \1_{\{ 0< z(T) \leq C_2  \}} \label{def_vrmv_xT}
\end{align}
where $C_1$ and $C_2$ are defined as,
\begin{align}
C_1\triangleq \max \left\{\frac{\rho^*}{\eta^*}, ~K_0^{-1}(1 -\gamma) \right\},~
C_2\triangleq \min \left\{\frac{\rho^*+2\beta^*}{\eta},~ K_0^{-1}(1 -\gamma) \right\},\label{def_vrmv_c1c2}
\end{align}
and $\rho^*$ and $\eta^*>0$ are the solution of the following two equations,
\begin{align*}
&\E \Big[ \left( \frac{\rho^*-\eta^* z(T)}{2} \right)z(T)\1_{  \left\{ K_0^{-1}(1-\gamma) < z(T) \leq C_1  \right\} } \Big]
- \beta^*\E\Big[ z(T) \1_{  \left\{ C_2 < z(T) \leq  K_0^{-1}(1-\gamma) \right\} } \Big]  \\
&~~~~~~~~~~~~~~~~~~~~~~~~+ \E\Big[ \left( \frac{\rho^*-\eta^* z(T)}{2} \right) z(T)\1_{ \left\{ 0 < z(T) \leq C_2 \right\} } \Big]=x_0, \\
&\E \Big[ \left( \frac{\rho^*-\eta^* z(T)}{2} \right) \1_{  \left\{ K_0^{-1}(1-\gamma) < z(T) \leq C_1  \right\} } \Big]
- \beta^*\E\Big[ \1_{  \left\{ C_2 < z(T) \leq  K_0^{-1}(1-\gamma) \right\} } \Big]  \\
&~~~~~~~~~~~~~~~~~~~~~~~~+ \E\Big[ \left( \frac{\rho^*-\eta^* z(T)}{2} \right) \1_{ \left\{ 0 < z(T) \leq C_2 \right\} } \Big]=x_d,
\end{align*}
and $\beta^*$ can be characterized by,
\begin{align*}
\beta^*=&\arg\min_{\beta \in[\underline{\beta},0]}~\Big\{ \E\Big[  \frac{ (\rho^*-\eta^* z(T))^2}{4} \1_{  \left\{ K_0^{-1}(1-\gamma) < z(T) \leq C_1  \right\} } \Big] + \beta^2 \E\Big[ \1_{  \left\{ C_2 < z(T) \leq  K_0^{-1}(1-\gamma) \right\} } \Big]  \\
&~~~~~~~~~~~~~~ + \E\Big[ \Big( \frac{ (\rho^*-\eta^* z(T))^2}{4} \Big) \1_{\{0 < z(T) \leq C_2\}} \Big] + \omega_{\vrmv}\beta \Big\}.
\end{align*}
\end{theorem}

\begin{figure}[!h]
\centering
\begin{subfigure}{0.45\textwidth}
   \includegraphics[width=\textwidth]{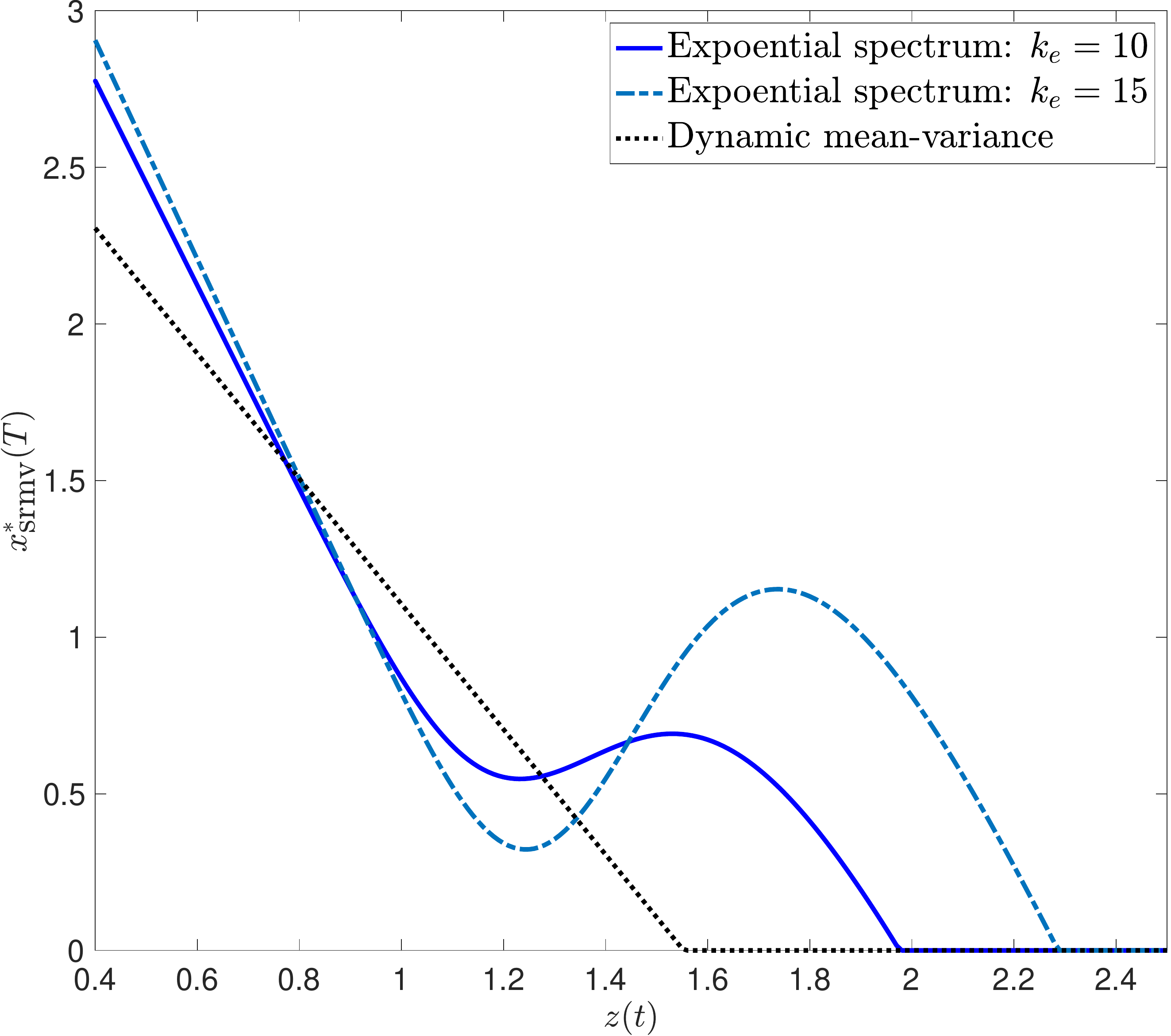}
   \caption{$x^*(T)$: model ($\cP_{\srmv}$) with exponential spectrum}
   \label{fig:srm_exp_xTzT}
\end{subfigure}
\begin{subfigure}{0.45\textwidth}
   \includegraphics[width=\textwidth]{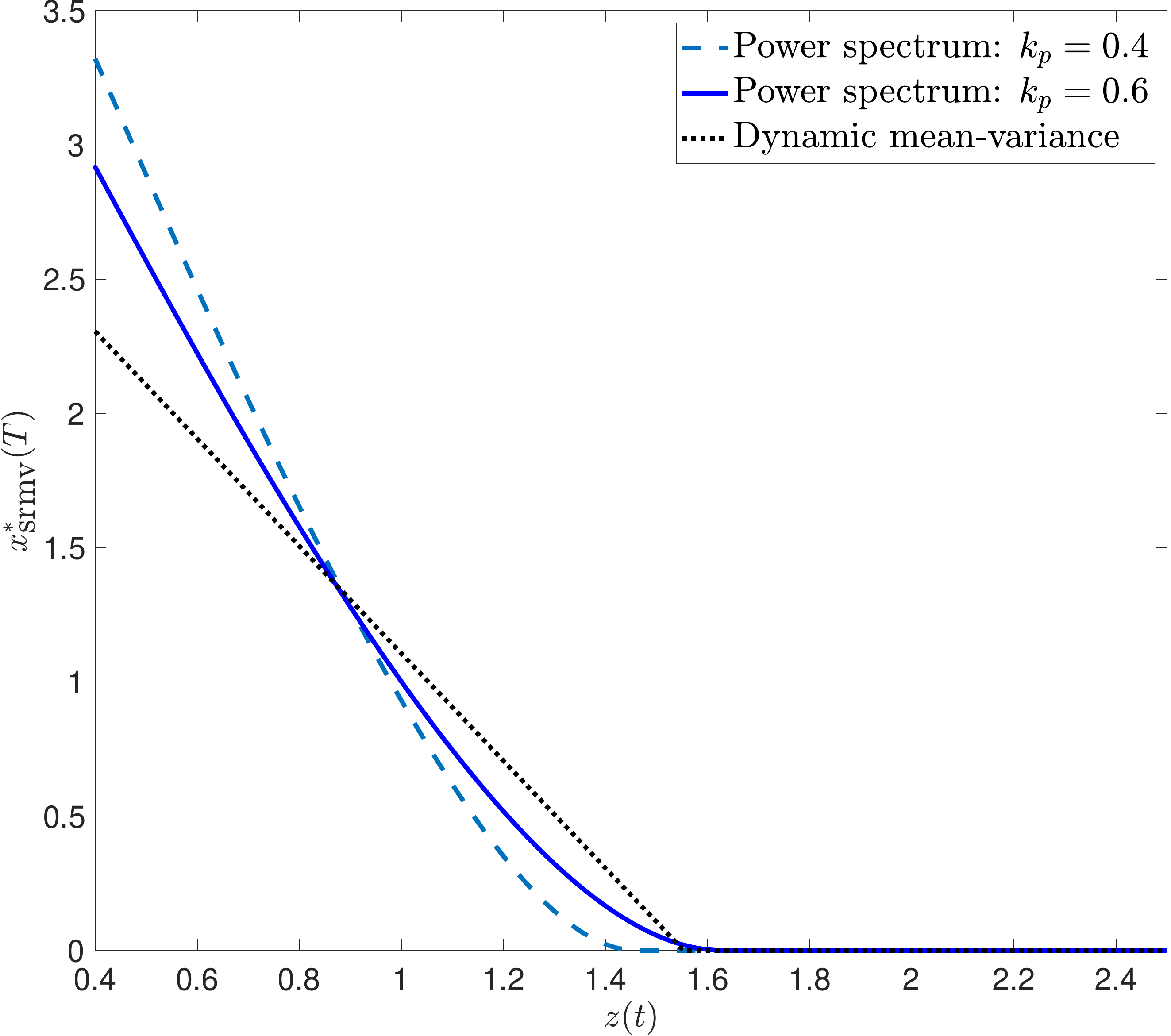}
   \caption{$x^*(T)$: model ($\cP_{\srmv}$) with power spectrum}
   \label{fig:srm_pow_xTzT}
\end{subfigure}
\begin{subfigure}{0.45\textwidth}
   \includegraphics[width=\textwidth]{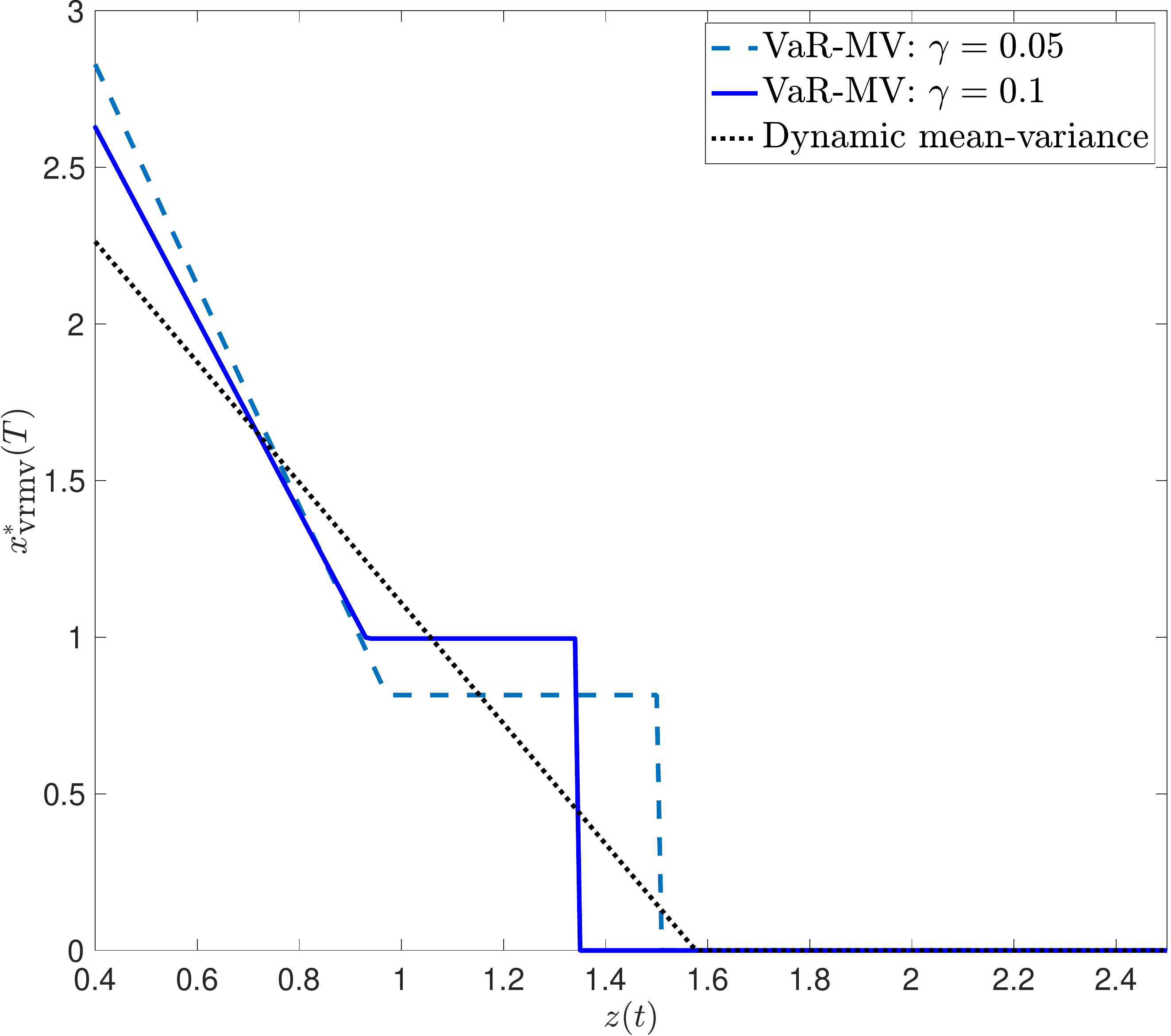}
   \caption{$x^*(T)$: model ($\cP_{\vrmv}$)}
   \label{fig:vrmv_xTzT}
\end{subfigure}
\caption{Terminal wealth $x^*(T)$ from models ($\cP_{\srmv}$) and ($\cP_{\vrmv}$) } \label{fig:xT_exp_pow_var}
\end{figure}

To examine the impacts of SRM and VaR in the hybrid portfolio models, Figure \ref{fig:xT_exp_pow_var} compares the terminal wealth resulted from model $(\cP_{\srmv})$ and model $(\cP_{\vrmv})$ with the one generated from pure dynamic MV model (e.g., Eq.(\ref{def_mv_xT})). The model parameters is from \cite{Berkelaar:2004} where a Black-Scholes type of market with one risky asset is considered. The parameters are $\mu(t)=0.1068$, $\sigma(t)=0.22$, and $r(t)=0.00408$ for $t\in [0,T=1]$ (year).
Figure \ref{fig:srm_exp_xTzT} and \ref{fig:srm_pow_xTzT} plot the wealth $x_{\srmv}^*(T)$ for the cases of the exponential spectrum (denoted by $x_{\srmv}^*(T)|_{\exp}$ with  $\omega_{\srmv}=0.5$) and the power spectrum (denoted by $x_{\srmv}^*(T)|_{\textrm{\pow}}$ with  $\omega_{\srmv}=1.5$) defined in Section \ref{sse:risk_measure}.  and Clearly, Eq.(\ref{def_mv_xT}) implies that the terminal wealth of the pure dynamic MV model is a piecewise linear function of $z(t)$ (denoted by $x_{\mv}^*(T)$ and indicated by the dotted line in all these figures. Figure \ref{fig:srm_exp_xTzT} shows that the exponential spectrum twists the terminal wealth $x^*_{\mv}(T)$, i.e., $x_{\srmv}^*(T)|_{\exp}$ is significantly higher than $x_{\mv}^*(T)$ in both good  market condition (in the region $z(T)<0.8$) and bad market condition ($z(T)>1.4$). However, $x_{\srmv}^*(T)|_{\exp}$ is in a lower position than the MV model in the mediate market condition (i.e., the region $z(T)\in(0.8,1.4)$). Similar pattern can be also observed from  portfolio model (see Figure \ref{fig:vrmv_xTzT}). Such a pattern of the wealth profile is desirable. First, a higher wealth level than $x^*_{\mv}(T)$ means {a higher} average profit in the good market condition ($z(T)<0.8$). Second, keeping the wealth above a positive level in {relatively} bad market condition (i.e. $z(T)\in (1,2)$) is known as the gambling strategy which may help to control the downside risk. Similar pattern is also reported in \cite{BasakShapiro:2001} in which the authors study the utility-based portfolio model combined with VaR constraint. As for the wealth profile of the power spectrum-based model ({Figure \ref{fig:srm_pow_xTzT}}), it behaves similar to $x^*_{\mv}(T)$, i.e., $x^*_{\srmv}(T)|_{\pow}$ only enhances the wealth level in the good market condition. In this sense, the SRM-MV model with {the exponential} spectrum and VaR-MV model are the more ideal models which better shape the terminal wealth.

\subsection{Optimal portfolio policy}\label{sse_policy}

This section focuses on determining the optimal portfolio policy for problems $(\cP_{\srmv})$ and $(\cP_{\vrmv})$. If we know the optimal terminal wealth $x^*(T)$, which is a random variable, we can find the portfolio policy $u^*(t)$ that generates the contingent claim $x^(T)$. To achieve this, we can treat the wealth process (\ref{def_xt}) as a backward stochastic differential equation (BSDE), where the unknown processes are $x^*(t)$ and $u^*(t)$, and the terminal condition is (\ref{def_srmv_xT}) or (\ref{def_vrmv_xT}). Although the solution to such a linear BSDE (i.e., the hedging policy) can be represented by the abstract martingale representation (as shown in Theorem 1.1 in \cite{KarouiBSDE:1997}) in a general market setting, our focus is on characterizing the structure of the solutions (or at least numerically) for specific markets. We consider three types of market models commonly used in academic research and practical applications: (i) the Black-Scholes market, where all market parameters have deterministic values, (ii) the market with mean-reverting returns, and (iii) the market with stochastic volatility.

\subsubsection{Black-Scholes Market}\label{sse_srmv_BS}

In this section, we consider the Black-Scholes type of market model, i.e., we impose the following assumption.
\begin{assumption} \label{asmp_BS}
All the market parameters, $r(t)$, $\{\mu_i(t)\}|_{i=1}^{n}$ and $\{\sigma_{ij}(t)\}|_{i=1,j=1}^{n,n}$ are deterministic functions of $t$ for all $t\in[0,T]$.
\end{assumption}

Under the Assumption \ref{asmp_BS}, we may {characterize} the closed-form solution of problems $(\cP_{\srmv})$ and $(\cP_{\vrmv})$. Recall the definition of the deflator process $z(t)$ in Eq. (\ref{def_zt}), which is equivalent to $\ln\left( \frac{z(T)}{z(t)}\right) $$=$$-\int_t^T \big( r(\tau)+\frac{1}{2}||\theta(\tau)||^2 \big )d\tau + \int_t^T \theta(\tau)^{\prime}dW(\tau)$.
The above expression and the Assumption \ref{asmp_BS} imply that, the random variable $\ln\big(z(T)/z(t)\big)$ follows a normal distribution $\mathcal{N}\big( m(t),v^2(t) \big)$ where the associated mean and variance are
\begin{align}
m(t) =-\int_t^T \big(r(\tau)+\frac{1}{2}||\theta(\tau)||^2\big) d\tau~~ \textrm{and}~~ v^2(t) =\int_t^T ||\theta(\tau)||^2 d\tau \label{def_mtvt}
\end{align}
for $t\in[0,T]$, respectively. For the convenience of presentation, we introduce the following functions of $t\in[0,T]$, $A(t) \triangleq  e^{m(t)+\frac{v(t)^2}{2}}$ and $B(t) \triangleq  e^{2 m(t)+ 2 v(t)^2}$.

Note that, at time $t=0$, since $z(0)=1$, the mean and variance of $\ln(z(T))$ are $m(0)$ and $v^2(0)$, respectively. Since $\ln\big( z(T)/z(t)\big)$ follows log-normal distribution, we may get the closed-form expression of the expectation (\ref{def_xt_expectation}). As for the time-$t$ portfolio policy $u^*(t)$, it can be computed by (see e.g., \cite{KaratzasShreve:1998}) the following equation,
\begin{align}
u^*(t)= -\big( \sigma(t)\sigma^{\top}(t)\big)^{-1} b(t) z(t) \frac{\partial x^*(t)}{\partial z(t)}. \label{BS_ut}
\end{align}
for $t\in [0,T]$.

We first focus on the model ($\cP_{\srmv}$). Substituting (\ref{def_srmv_xT}) to (\ref{def_xt_expectation}) yields,\footnote{Detail on the computation (\ref{def_srmv_xt_BS}) is provided in {\ref{appendix_proof_srmv_vrmv_xt_BS}}.}
\begin{align}
x_{\srmv}^*(t) &= \frac{\rho^*}{2}A(t)\Phi\big( \kappa_1(t) \big) - \frac{\eta^* z(t)}{2}B(t)\Phi\big( \kappa_2(t)\big)\notag\\
       &+\frac{\omega_{\srmv}}{2} \int^{\kappa_1(t)+v(t)}_{-\infty} e^{ s \cdot v(t)+ m(t)} \psi\left(1-K_0\big( z(t)e^{ s \cdot v(t)+ m(t)}\big) \right)\phi(s)ds, \label{def_srmv_xt_BS}
\end{align}
where $\kappa_1(t)\triangleq\frac{\ln(z^{\dag}/z(t)) - m(t)}{v(t)}-v(t)$, $\kappa_2(t)\triangleq\kappa_1(t)-v(t)$ and $K_0(y) = \Phi\left(\frac{\ln(y)-m(0)}{v(0)} \right)$.

In Theorem \ref{thm_srmv_xT}, we need solve systems of three equations (\ref{equ_zdag_sr}), (\ref{equ_xd_sr}) and (\ref{equ_x0_sr}) to compute the unknown parameters $s^{\dag}$, $\rho^*$ and $\eta^*$. Under the Assumption \ref{asmp_BS}, we may write out the closed-form expression of these equations. Indeed, the equation (\ref{equ_srmv_s_dag}) can be written as
\begin{align}
\rho^* - \eta^* e^{m(0) + v(0)\Phi^{-1}(1-s^{\dag})} + \omega_{\srmv}\psi(s^{\dag}) = 0.\label{equ_zdag_sr_BS}
\end{align}
The equations (\ref{equ_x0_sr}) and (\ref{equ_xd_sr}) can be reformulated as \footnote{The right hand side of equation (\ref{equ_x0_sr}) is just the equation (\ref{def_srmv_xt_BS}) by setting $t=0$.}
\begin{align}
x_0 &= \frac{\rho^*}{2}A(0)\Phi( \kappa_1(0))-\frac{\eta^* }{2}B(0)\Phi( \kappa_2(0))\notag\\
       &~~+\frac{\omega_{\srmv}}{2} \int^{\kappa_1(0)+v(0)}_{-\infty} e^{ s \cdot v(0)+ m(0)} \psi\left(1 - K_0\big( e^{ s \cdot v(0)+ m(0)}\big) \right)\phi(s)ds, \label{equ_x0_sr_BS}\\
x_d &= \frac{\rho^*}{2} \Phi( \kappa_1(0)+v(0)) -\frac{ \eta^* }{2} A(0) \Phi( \kappa_1(0))\notag\\
    &~~~+\frac{\omega_{\srmv}}{2} \int^{ \kappa_1(0)+v(0)}_{-\infty} \psi\left( 1 - K_0\big(e^{s \cdot v(0) +m(0)}\big) \right) \phi(s)ds.\label{equ_xd_sr_BS}
\end{align}
Combining Eq. (\ref{BS_ut}) with Eq. (\ref{def_srmv_xt_BS}), we have the optimal portfolio policy $u_{\srmv}^*(t)$ as,
\begin{align}
u_{\srmv}^*(t)&=u^1_{\srmv}(t) + \omega_{\srmv} \cdot u^2_{\srmv}(t), \label{def_ut_srmv_BS}\\
u^1_{\srmv}(t)&=\frac{1}{2}\big(\sigma(t)\sigma^{\top}(t)\big)^{-1}b(t)
\Big(  \frac{\rho^* A(t) \phi(\kappa_1(t))}{ v(t)}
+ \eta^* B(t) z(t) \Big( \Phi\big( \kappa_2(t) \big)- \frac{\phi\big( \kappa_2(t)\big) }{v(t)}\Big)\Big), \notag \\
u^2_{\srmv}(t)&= \frac{1}{2}\big(\sigma(t)\sigma^{\top}(t)\big)^{-1}b(t)\Big(\frac{z^{\dag}}{v(t)z(t)} \psi(1-K_0(z^{\dag}))\phi(\kappa_1(t)+v(t))\notag\\
&+\int^{\kappa_1(t)+v(t)}_{-\infty} z(t)e^{2(s v(t)+m(t))} \psi^{\prime}(1-K_0(z(t)e^{sv(t)+m(t)}) )K_0^{\prime}(z(t)e^{sv(t)+m(t)})\phi(s)ds\Big),\notag
\end{align}
where $\psi^{\prime}(\cdot)$ and $K_0^{\prime}(\cdot)$ are the first-order derivatives of functions $\psi(\cdot)$ and $K_0(\cdot)$ respectively. Following the decomposition of the terminal wealth (\ref{def_srmv_xT_decompose}), SRM-MV hybrid policy (\ref{def_ut_srmv_BS}) also can be also decomposed as a weighted summation of the dynamic MV portfolio policy $u_{\srmv}^1(t)$ and the SRM spectrum related policy
$u^2_{\srmv}(t)$.

We then turn to the VaR-MV hybrid model $(\cP_{\vrmv})$. To express the solution in a more compact formulation, we first introduce three functions of $z(t)$ for any $t\in [0,T]$ as $\iota_1(t) $$\triangleq$$ \frac{\ln\big( (C_2)^+/z(t)\big)-m(t)}{v(t)}-2v(t)$, $\iota_2(t) $$\triangleq$$\frac{\Phi^{-1}(1-\gamma)\cdot v(0)+m(0)-\ln z(t)-m(t)}{v(t)}-2v(t)$, and $\iota_3(t) \triangleq \frac{\ln\big( (C_1)^+/z(t)\big) - m(t)}{v(t)}-2v(t)$,
for $t\in [0,T]$. Then, we may compute the optimal wealth process $x^*_{\vrmv}(t)$ and portfolio policy $u^*_{\vrmv}(t)$ by (\ref{def_xt_expectation}), (\ref{def_vrmv_xT}) and (\ref{BS_ut}), i.e., it has
\begin{align}
x_{\vrmv}^*(t)&=  A(t)\Big( \frac{\rho^*}{2}\Phi\big(\iota_3(t)+v(t)\big) - \big(\frac{\rho^*}{2}+\beta^*\big)\big(\Phi\big(\iota_2(t)+v(t)\big)-\Phi\big(\iota_1(t)+v(t)\big)\big) \Big)\notag\\
&~~~- \frac{\eta^* z(t)B(t)}{2} \Big( \Phi\big(\iota_3(t) \big) -\Phi\big(\iota_2(t)\big) + \Phi\big( \iota_1(t) \big) \Big), \label{vrmv_bs_xt} \\
u_{\vrmv}^*(t)&=(\sigma(t)\sigma^{\top}(t))^{-1}b(t)\Big( \frac{A(t)}{v(t)}\Big( \frac{\rho^*}{2}\phi\big(\iota_3(t)+v(t) \big) - \big(\frac{\rho^*}{2} +\beta^*\big)\big(\phi\big( \iota_2(t)+v(t) \big)\notag\\
&~~- \phi\big(\iota_1(t) +v(t)\big)\big) \Big) + \frac{\eta^* z(t)B(t)}{2}\Big( \Phi\big(\iota_3(t) \big)-\Phi\big(\iota_2(t) \big)+\Phi\big(\iota_1(t) \big) \notag\\
&~~- \frac{1}{v(t)}\big(\phi\big(\iota_3(t) \big) - \phi\big(\iota_2(t) \big) + \phi\big(\iota_1(t) \big)\big) \Big) \Big), \label{vrmv_bs_ut}
\end{align}
where the Lagrangian multipliers $\rho^*$ and $\eta^*>0$ are the solution of the following equations,
\begin{align}
x_d =& \frac{\rho^*}{2}\Phi\big( \iota_3(0)+2v(0) \big) - \big( \frac{\rho^*}{2}+\beta^*\big) \big(\Phi\big( \iota_2(0) + 2v(0)\big)
-\Phi\big( \iota_1(0)+2v(0) \big)\big) \notag \\
& - \frac{\eta^*}{2} A(0) \Big( \Phi( \iota_3(0)+v(0) )-\Phi(\iota_2(0)+v(0) )+\Phi(\iota_1(0)+v(0) ) \Big), \label{vrmv_equ_x0_BS} \\
x_0 =& A(0)\Big( \frac{\rho^*}{2}\Phi\big( \iota_3(0) +v(0)\big) - \big( \frac{\rho^*}{2} + \beta^*\big)\big( \Phi(\iota_2(0)+v(0) ) -\Phi(\iota_1(0)+v(0) )\big) \Big) \notag \\
& - \frac{\eta^* B(0)}{2} \Big( \Phi(\iota_3(0) ) -\Phi(\iota_2(0) )+\Phi(\iota_1(0) ) \Big).\label{vrmv_equ_xd_BS}
\end{align}
and the parameter $\beta^*$ can be characterized by,
\begin{align*}
\beta^*=&\arg\min_{\beta\in[\underline{\beta},0]}~\Big\{-\frac{\rho^*\eta^*}{2}A(0)
\Big( \Phi\big( \iota_3(0) +v(0) ) - \Phi\big( \iota_2(0) +v(0)\big)+\Phi\big( \iota_1(0) +v(0)\big)   \Big) \\
&~~~~~~~~~~~~~~ +\frac{(\eta^*)^2}{4}B(0)\Big( \Phi\big( \iota_3(0) \big) - \Phi\big( \iota_2(0)\big) + \Phi\big( \iota_1(0)\big)   \Big) \\
&~~~~~~~~~~~~~~ +\frac{(\rho^*)^2}{4}\Big( \Phi\big(\iota_3(0)+2v(0) \big)-\Phi\big(\iota_2(0)+2v(0)\big)+\Phi\big(\iota_1(0)+2v(0)\big)   \Big)\\
&~~~~~~~~~~~~~~ +\beta^2\Big( \Phi\big(\iota_2(0)+2v(0)\big)-\Phi\big(\iota_1(0)+2v(0)\big)   \Big) + \omega_{\vrmv}\beta \Big\}.
\end{align*}
{The detail of} deriving equations (\ref{vrmv_bs_xt}) and (\ref{vrmv_bs_ut}) are given in {\ref{appendix_proof_srmv_vrmv_xt_BS}}.

Next, we present a numerical example to examine the characteristics of the SRM-MV and VaR-MV hybrid portfolio policies. As a benchmark, we use the dynamic MV portfolio optimization model (with no bankruptcy restrictions, $x(T)\geq 0$). Using Eqs. (\ref{def_mv_xT}), (\ref{def_xt_expectation}), and (\ref{BS_ut}), we can express the dynamic MV portfolio policy (referred to as $u^*_{\mv}(t)$) as follows: \footnote{In the literature, \cite{BieleckiJinPliskaZhou:2005} provides the solution of the dynamic MV portfolio optimization model with no-bankruptcy constraint. However, in their model, the policy is represented by a {fictitious} asset. Different from their solution, we express the MV policy as a function of $z(t)$. The detail is provided in the Appendix.}
\begin{align}
u_{\mv}^*(t) &=  \frac{1}{2}\big( \sigma(t)\sigma^{\top}(t)\big)^{-1}b(t) \Big( \frac{\rho_{\mv}A(t)\phi( \kappa_1(t) )}{v(t)}\notag\\
&~~~~~~~~~~~~~~~~~+{\eta_{mv}} B(t)z(t)\Big( \Phi(\kappa_1(t)- v(t) ) -\frac{\phi( \kappa_1(t)-v(t))}{v(t)}  \Big)
   \Big), \label{def_u_mv}
\end{align}
where $\eta_{\mv}$ and $\rho_{\mv}$ are the solution of the equations (\ref{equ_x0_sr_BS}) and (\ref{equ_xd_sr_BS}) when $\omega_{\srmv}=0$.

\begin{figure}[!h]
\centering
\begin{subfigure}{0.47\textwidth}
   \includegraphics[width=\textwidth]{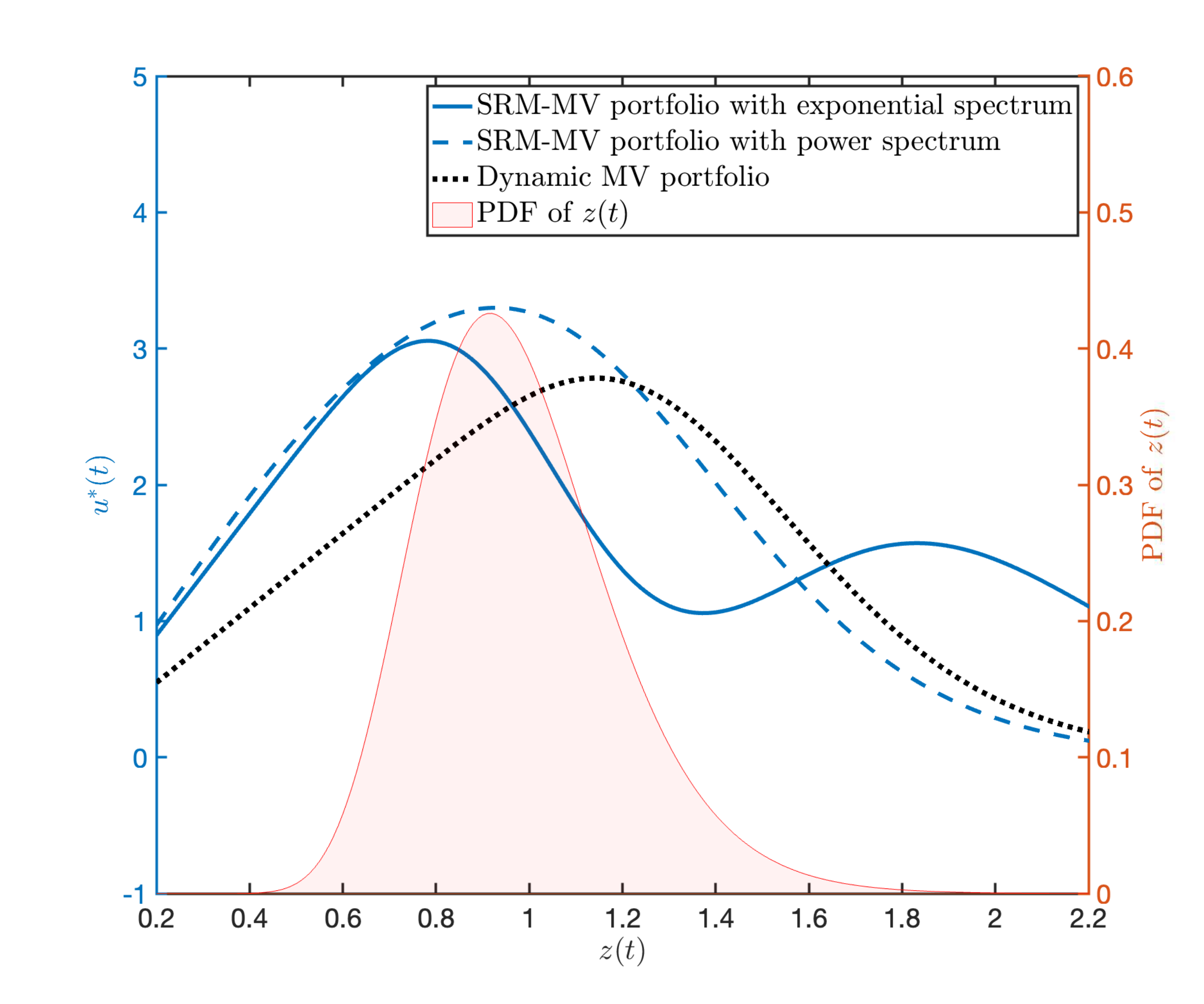}
   \caption{SRM-MV portfolio: $u_{\srmv}^*(t)$ and $z(t)$  }
   \label{fig:srmv_ztut}
\end{subfigure}
\begin{subfigure}{0.47\textwidth}
   \includegraphics[width=\textwidth]{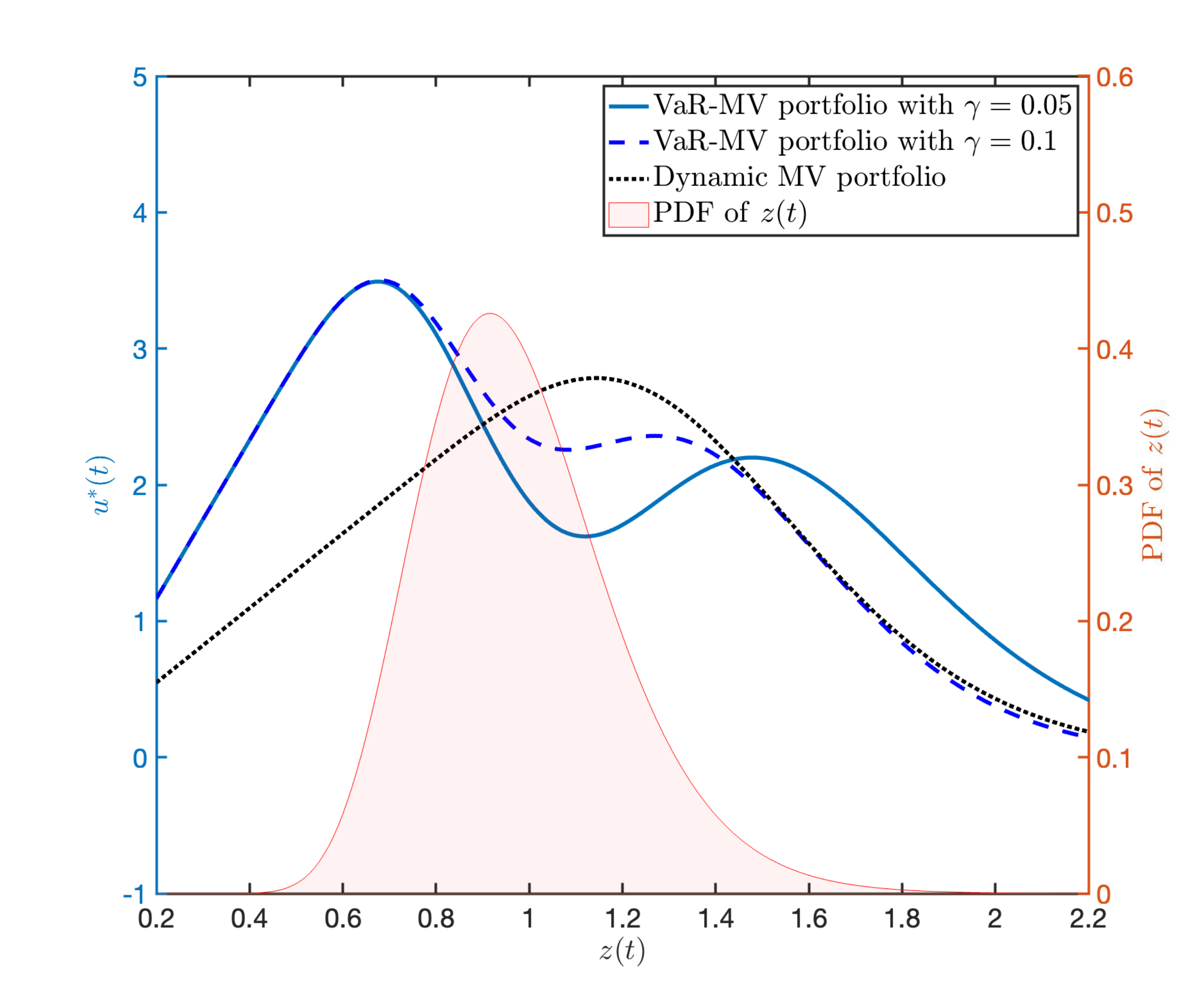}
   \caption{VaR-MV portfolio: $u_{\vrmv}^*(t)$ and $z(t)$ }
   \label{fig:vrmv_ztut}
\end{subfigure}
\caption{Portfolios $u^*_{\srmv}$ and $u^*_{\vrmv}$ in response to SPD $z(t)$} \label{fig:srmv_vrmv_ztut}
\end{figure}

We used the same parameters as shown in Figure \ref{fig:xT_exp_pow_var}, specifically, $\mu(t)=0.1068$, $\sigma(t)=0.22$, and $r(t)=0.00408$ for $t\in [0,T]$ with $T=1$. We also set $k_e=10$ and $k_p=0.6$. For the weighting parameters, we used $\omega_{\srmv}=0.5$ for the exponential spectrum-based model, $\omega_{\srmv}=1.5$ for the power spectrum-based model, and $\omega_{\vrmv}=0.8$. Figure \ref{fig:srmv_ztut} plots the portfolio policies $u_{\srmv}^*(t)$ and $u_{\vrmv}^*(t)$ as a function of the SPD $z(t)$ at intermediate time $t=T/2$. To distinguish different portfolio policies, we use $u^*_{\srmv}(t)|_{\exp}$ and $u^*_{\srmv}(t)|_{\pow}$ to denote the SRM-MV hybrid portfolio policies resulting from the exponential spectrum and the power spectrum, respectively. In Figure \ref{fig:srmv_ztut}, the policies $u^*_{\srmv}(t)|_{\exp}$, $u^*_{\srmv}(t)|_{\pow}$ and $u^*_{\mv}(t)$ are indicated by the solid line, the dashed line and the dashed line, respectively. Since the SPD $z(t)$ is a random variable, we also plot the probability density function of $z(t)$ in the second Y-axis. Then, the shaded area indicates the probability distribution of the $z(t)$. Consistent with the wealth profile of these models (e.g., see Figure \ref{fig:srm_exp_xTzT}), the policy $u_{\srmv}^*(t)|_{\exp}$ behaves differently from the benchmark policy $u^*_{\mv}(t)$. In both good market condition ($z(t)<0.8$) and bad market condition ($z(T)>1.7$), $u_{\srmv}^*(t)|_{\exp}$ has a much higher position than $u^*_{\mv}(t)$. However, in the moderate market condition (e.g., $z(t)\in(0.8,1.7)$), $u_{\srmv}^*(t)|_{\exp}$ is significantly lower than $u^*_{\mv}(t)$. On the other hand, the power spectrum-based portfolio $u^*_{\srmv}(t)|_{\pow}$ does not seem to have a similar pattern as the exponential spectrum-based counterpart. It appears more likely to the MV portfolio $u^*_{\mv}(t)$, which only has a single peak in the region $z(t)\in[0.4,0.8]$. Figure \ref{fig:vrmv_ztut} compares the VaR-MV portfolio $u^*_{\vrmv}(t)$ with the MV portfolio $u^*_{\mv}(t)$ when the confidence level is $\gamma=0.05$ and $\gamma=0.1$. We can observe that the basic pattern of $u^*_{\vrmv}(t)$ is almost as same as $u^*_{\srmv}(t)|_{\exp}$.

\begin{figure}[!h]
\centering
\begin{subfigure}{0.47\textwidth}
   \includegraphics[width=\textwidth]{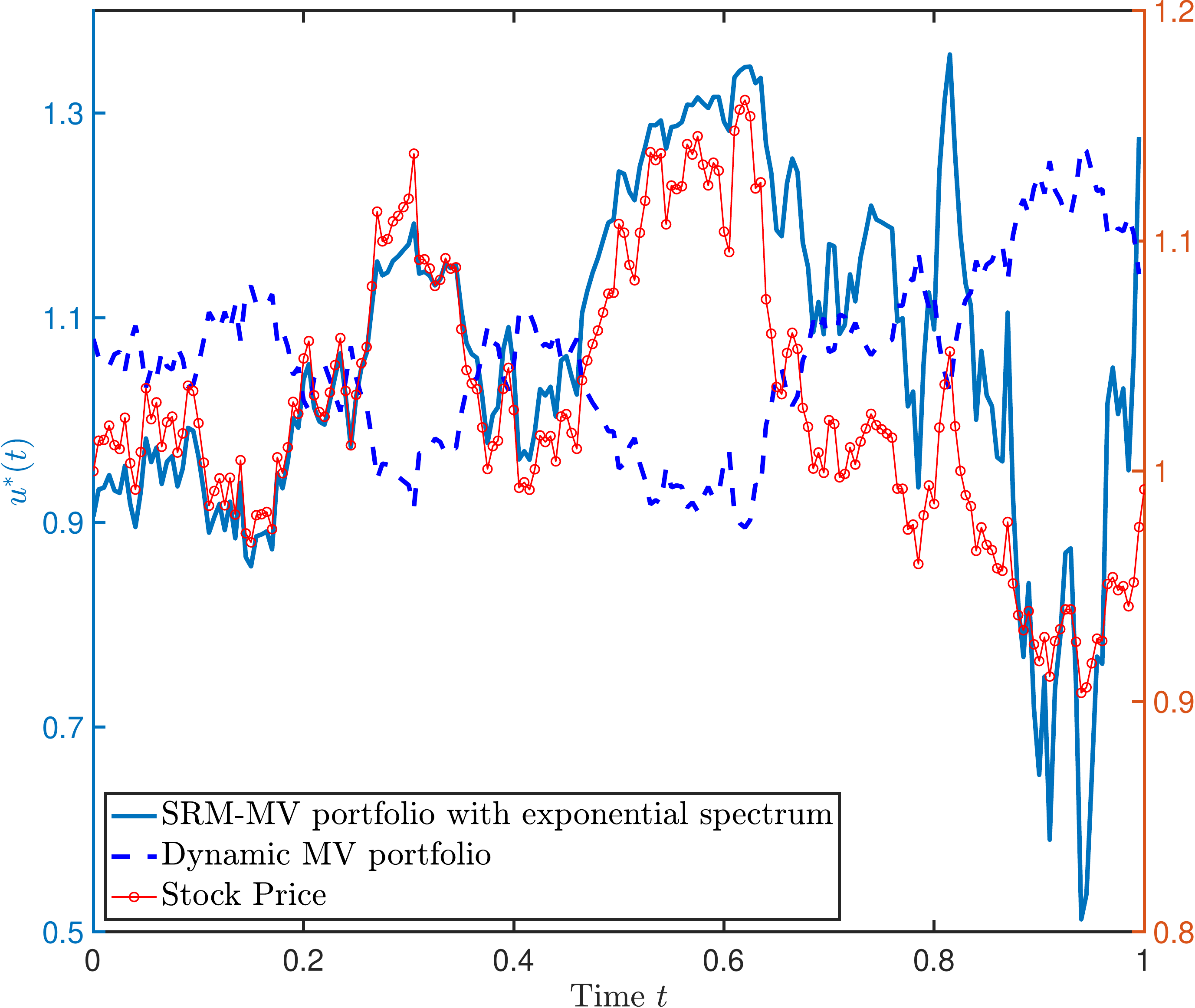}
   \caption{Model $(\cP_{\srmv})$: $u^*(t)$ and  $z(t)$ }
   \label{fig:srmv_path_Stut}
\end{subfigure}
\begin{subfigure}{0.47\textwidth}
   \includegraphics[width=\textwidth]{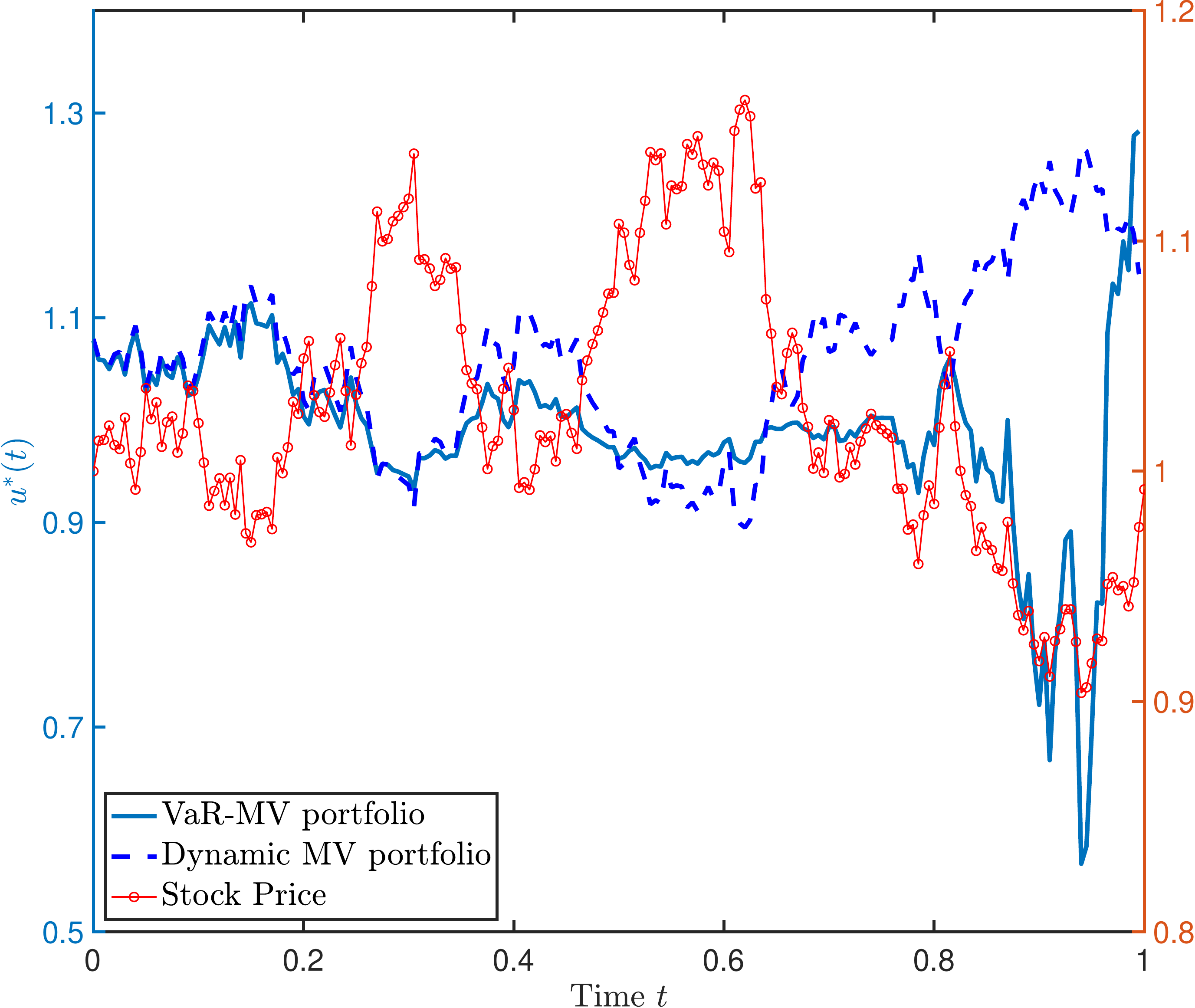}
   \caption{VaR-MV model: $u^*(t)$ and $z(t)$}
   \label{fig:vrmv_path_Stut}
\end{subfigure}
\caption{Portfolio $u^*(t)$ in response to $z(t)$} \label{fig:path_Stut}
\end{figure}

While the previous figures compared different portfolios as a function of the market state at a fixed time, we will now examine the time series of various portfolio policies. A straightforward approach is to observe how the portfolio reacts to changes in the stock price.\footnote{In Black-Scholes model, the asset price $S(t)$ and SPD $z(t)$ are a one-to-one mapping for a fixed sample path. In this example, as $\mu(t)=\mu$, $r(t)=r$ and $\sigma(t)=\sigma$ for $t\in[0,T]$, from (\ref{def_St}) and (\ref{def_zt}), it has $\ln(S(t))=\left(\mu-\frac{\sigma^2}{2} -\frac{r\sigma}{\theta} -\frac{\theta \sigma}{2}\right)t -\frac{\sigma}{\theta}\ln(z(t))$ when $S_0=1$.}  In this test, we only consider the exponential spectrum based model. Figure \ref{fig:path_Stut} illustrates the portfolio allocations $u^*_{\srmv}(t)|_{\exp}$, $u^*_{\vrmv}(t)$, and $u^*_{\mv}(t)$ against the stock price $S(t)$ resulting from one simulation path of the stock price (the red line with values marked by the second Y-axis in Figure \ref{fig:path_Stut}). It is apparent that the MV portfolio $u^*_{\mv}(t)$ and the stock price $S(t)$ are strongly negatively correlated, as the investment target minimizes the variance of the cumulative wealth. However, the hybrid portfolio policies $u^*_{\srmv}(t)|_{\exp}$ and $u^*_{\vrmv}(t)$ are noticeably different from the MV policy. Specifically, $u^*_{\srmv}(t)|_{\exp}$ displays a certain degree of trend-following behavior. This pattern may help to increase gains during uptrends and mitigate losses during downtrends. Interestingly, the VaR-MV policy exhibits an asymmetrical pattern, where $u^*_{\vrmv}(t)$ behaves like the MV policy (negative-feedback trading) when the stock price goes up, and behaves like the SRM-MV policy (trend-following) when the stock price goes down. Such a pattern helps to control both the variance and the downside risk.

\subsubsection{Market with mean-reverting return and stochastic volatility} \label{subsubsec_stochastic}
In this section, we examine the case when the market parameters are stochastic processes. We focus on one commonly used cases in academic study, namely, the market with mean-reverting return (\citealp{Wachter:2002,KimOmberg:1996}). We adopt a similar setting in \cite{Wachter:2002}, i.e., this simple market has one risky asset and a risk-free asset. The risk-free rate is a constant $r(t)= r\geq 0$ for $t\in [0,T]$ and the price process of risky asset is $dS(t)=S(t)\Big(\mu(t)dt + \sigma dW(t)\Big)$ with a given $S(0)=s_0$ where $\sigma>0$ is the constant volatility, $\mu(t) \triangleq \theta(t) \sigma + r$ is the return rate and $\theta(t)$ is the instantaneous Sharpe ratio which satisfies the following Ornstein-Uhlenbeck (OU) process, $d\theta(t)=\lambda\big( \bar{\theta}-\theta(t)\big)dt+\gamma dW(t)$ where $\theta(0) = \theta_0$ and $\bar{\theta}$, $\lambda\geq 0$ and $\gamma\in\mathbb{R}$ are the constant parameters. As the return rate $\mu(t)$ is an affine function of $\theta(t)$, this setting implies that $\mu(t)$ satisfies OU process. \footnote{The return rate $\mu(t)$ satisfies $d\mu(t)=\lambda_{\mu}(\bar{\mu}-\mu(t))dt+\gamma_{\mu} dW(t)$ where $\lambda_{\mu}=\lambda$, $\gamma_{\mu}= \sigma \gamma$ and $\bar{\mu}=\bar{\theta}\sigma+r$. }

Under the above setting, the deflator process $z(t)$ and the optimal wealth $x^*(T)$ still take the similar form as in ($\ref{def_zt}$) and (\ref{def_srmv_xT}) or (\ref{def_vrmv_xT}), respectively. However, as the market state $\theta(t)$ follows the OU process, the random variable $z(T)/z(t)$ does not follow Log-Normal distribution any more.\footnote{Indeed, there does not exist the closed-form expression of the probability density function of $z(T)$. } That is to say, we can not compute the expectation (\ref{def_xt_expectation}) analytically any more. As for the numerical method, we may adopt the Monte Carlo-based method to compute the expectation in (\ref{def_xt_expectation}). Specifically, at any time $t\in[0,T)$, given the state variables (i.e., $z(t)$), we generate the sample paths of the $z(\tau)$ for $\tau\in [t,T]$. For any sample of $z(T)$, we then compute the correspondent sample of $x^*(T)$ by (\ref{def_srmv_xT}) and further compute the sample average as an approximation of (\ref{def_xt_expectation}). Besides the Monte Carlo method, we may characterize $x^*(t)$ by solving the partial differential equation (PDE). This method is based on the Feynman-Kac formula (see, e.g., \cite{Pham:2009} and \cite{YongZhou:1999}), i.e., computing the conditional expectation (\ref{def_xt_expectation}) is equivalent to solving the associated PDE. Since $x^*(t)$ is related to two state variables $z(t)$ and $\theta(t)$ at time $t$, we use $X(t,z,\theta)$ to denote the optimal wealth process $x^*(t)$.\footnote{When there is no ambiguity, we ignore the argument $t$ in $z(t)$ and $\theta(t)$ to simplify the notations.} It can be verified that $X(t,z,\theta)$ satisfies the following PDE (see \cite{Wachter:2002,GaoLiYao:2018}),
\begin{align}
&\frac{\partial X}{\partial t} + z(\theta^2-r)\frac{\partial X}{\partial z} + \Big(\lambda\bar{\theta}-(\lambda+\gamma)\theta\Big)\frac{\partial X}{\partial \theta} + \frac{1}{2}\theta^2z^2\frac{\partial^2 X}{\partial z^2} + \frac{1}{2}\gamma^2\frac{\partial^2 X}{\partial \theta^2} - \gamma z \theta\frac{\partial^2 X}{\partial \theta \partial z} = rX, \label{PDE_H}
\end{align}
where the terminal condition is $X(T,z,\theta)= x_{\srmv}^*(T)$ or $x_{\vrmv}^*(T)$ (i.e., Eq. (\ref{def_srmv_xT}) or Eq.(\ref{def_vrmv_xT})). Moreover, the optimal portfolio policy can be computed as
\begin{align}
u^*(t) = \frac{1}{\sigma}\Big(-z\theta\frac{\partial X }{\partial z} + \gamma\frac{\partial X }{\partial \theta}\Big),~~\textrm{for any}~~t\in[0,T]. \label{PDE_u}
\end{align}
Although the PDE (\ref{PDE_H}) is similar to the one given in \cite{Wachter:2002} or \cite{GaoLiYao:2018}, due to the terminal condition, it does admit a closed form solution. Thus, we need to use the numerical method to solve the PDE (\ref{PDE_H}). Once we achieve the optimal wealth process $X(t,z,\theta)$, we can derive the optimal portfolio policy $u_{\srmv}^*(t)$ or $u_{\vrmv}^*(t)$ from the formula (\ref{PDE_u}).
We then another popular market setting, which is also known as the Heston's model (\cite{Kraft:2005}), models the stochastic volatility. In this model, the price of risky asset $S(t)\in \mathbb{R}$ follows the following process: $dS(t) = S(t)\Big(\mu dt + \sqrt{\nu(t)}dW(t)\Big)$ and $\nu(t) = \iota (\bar{\nu}-\nu(t))dt + \xi\sqrt{\nu(t)}d\hat{W}(t)$, where $\hat{W}(t)$ is standard Brownian motion, $\nu(t)$ is the instantaneous variance, $\bar{\nu}>0$ is the long-run average variance of the price, $\iota>0$ is the rate at which $\nu(t)$ reverts to $\bar{\nu}$ and $\xi>0$ is the volatility of the volatility which determines the variance of $\nu(t)$. Similarly, we may use the PDE approach to characterize optimal wealth. We use $X(t,z,\nu)$ to denote the optimal wealth process $x^*(t)$ at time $t$. Then $X(t,z,\nu)$ satisfies the following PDE,
\begin{align}
z\Big(\frac{(\mu-r)^2}{\nu}-r\Big)\frac{\partial X}{\partial z} &+ \Big( {\iota(\bar{\nu}-\nu)}-(\mu-r)\xi \Big)\frac{\partial X}{\partial \nu} + \frac{\partial X}{\partial t} + \frac{1}{2}\frac{z^2(\mu-r)^2}{\nu}\frac{\partial^2 X}{\partial z^2} \notag \\
& + \frac{1}{2}(\xi^2\nu)\frac{\partial^2 x}{\partial \nu^2} - z(\mu-r)\xi\frac{\partial^2 X}{\partial z \partial \nu} = r X \label{thm_PDE_F}
\end{align}
where the terminal condition $X(T,z,\nu)$ =  $x_{\srmv}^*(T)$ or $x_{\vrmv}^*(T)$. Moreover, the optimal portfolio policy can be computed by,
\begin{align*}
u^*(t) = \frac{1}{\sqrt{\nu}}\Big( -\frac{z(\mu-r)}{\sqrt{\nu}}\frac{\partial X }{\partial z} + \xi\sqrt{\nu}\frac{\partial X  }{\partial \nu} \Big) = -\frac{z(\mu-r)}{\nu}\frac{\partial X }{\partial z} + \xi\frac{\partial X }{\partial \nu}.
\end{align*}
Besides the above two special cases, \cite{Duffie:Economica2003} have shown that the pricing problem of the contingent claim can be solved semi-analytically by the inverse Fourier Transformation for the market modeled by the affine jump-defusions. That is to say, the optimal wealth $x_{\srmv}^*(t)$ or $x_{\vrmv}^*(t)$ can be also computed numerically by their method for a more general market setting.

\section{Performance Analysis} \label{sec_example}

In this section, we present numerical experiments that evaluate the performance of the proposed dynamic and static hybrid portfolio optimization models. To do this, we apply these models to a practical scenario of constructing a pension fund comprising of three risky assets: the S$\&$P 500 index (SPI), the Emerging Market Index (EMI), and the Small Capital Index (SCI). To calibrate the basic annual statistics of these assets, we utilize historical data provided by \cite{CuiLiLi:2017}. The expected value and covariance matrix of the annual returns are presented in Table \ref{table_3assets_data}, while the risk-free return rate is based on a long-term bond at $3\%$. Based on this data, we construct the continuous-time market model by assuming that the asset prices follow the Multidimensional Geometrical Brownian Motion (GBM) with constant parameters. Specifically, we set $n=3$, $\mu(t)=\mu$, and $\sigma(t)=\sigma$ for $t\in[0,T]$ in the asset price model (\ref{def_St}). We use $R_i(t)=S_i(t)/S_i(0)$ to represent the total return of the $i$-th asset in the horizon $[0, t]$. As the price $S_i(t)$ is generated from the GBM, the log-return follows the Log-normal distribution, i.e., $ \ln( R_i(t)) \sim \mathcal{N}\big( \E[R_i(t)], \cV[R_i(t)]\big)$, where
\begin{align}
\E[R(t)] = t\big(\mu_i -\frac{1}{2}\sum_{j=1}^n\sigma_{i,j}^2\big)~~\textrm{and}~~\cV[R(t)] = t\sum_{j=1}^n \sigma_{i,j}^2. \label{Ln_ER_VR}
\end{align}
Using the Eq. (\ref{Ln_ER_VR}) and the annual statistics of the returns in Table \ref{table_3assets_data}, we may retrieve the parameters $\mu$ and $\sigma$ in the assets' price model (\ref{def_St}) as
\begin{align}
\mu=\begin{pmatrix}
0.1471\\
0.1566\\
0.1651\\
\end{pmatrix}~~~~
\sigma =
\begin{pmatrix}
0.1365 & 0.0568 & 0.0709\\
0.0568 & 0.2384 & 0.0898\\
0.0709 & 0.0898 & 0.1740
\end{pmatrix}.\label{data_mu_sigma}
\end{align}

\begin{table}[!h]
  \centering
  \begin{tabular}{c c c c}
      \toprule
                     & SPI & EMI & SCI   \\
      \cmidrule(l){2-4}
      Expected Return Rate & 0.121 & 0.130 & 0.151  \\
      \midrule
      \midrule
       Covariance Matrix & SPI & EMI & SCI    \\
       \cmidrule(l){1-1}    \cmidrule(l){2-4}
       SPI & 0.0342 & 0.0355 & 0.0351  \\
       EMI & 0.0522 & 0.0028 & 0.0540  \\
       SCI & 0.1645 & 0.0504 & 0.0576  \\
     \bottomrule
   \end{tabular}
  \caption{The statistics of the annual returns of the assets} \label{table_3assets_data}
\end{table}

To evaluate the effectiveness of different portfolio policies, we will generate $10,000$ sample paths of the asset prices using the model (\ref{def_St}). These price paths will then be used to implement various portfolio policies and record the resulting terminal wealth for each sample path. For the model $(\cP_{\srmv})$ and the model $(\cP_{\vrmv})$, we will use the results from Section \ref{sse_srmv_BS} to compute the dynamic policies. Specifically, we will solve Eqs. (\ref{equ_x0_sr_BS}) and (\ref{equ_xd_sr_BS}) for the SRM-MV policy (\ref{def_ut_srmv_BS}), and Eqs. (\ref{vrmv_equ_x0_BS}) and (\ref{vrmv_equ_xd_BS}) for the VaR-MV policy (\ref{vrmv_bs_ut}). For the static counterparts of these models $(\cP_{\srmv}^{\static})$ and $(\cP_{\vrmv}^{\static})$, we will use a discrete scenario approach outlined in the supplementary file to solve the associated mathematical programming problems for the policies.\footnote{The static SRM-MV problem ($\cP_{\srmv}^{\static}$) is written as a convex quadratic programming problem and the static VaR-MV problem ($\cP_{\vrmv}^{\static}$) is reformulated as a mixed-integer QP problem. All these problems are solved by calling commercial solver GUROBI \cite{gurobi}.}

\begin{figure}[!h]
\centering
\begin{subfigure}{0.49\textwidth}
   \includegraphics[width=\textwidth]{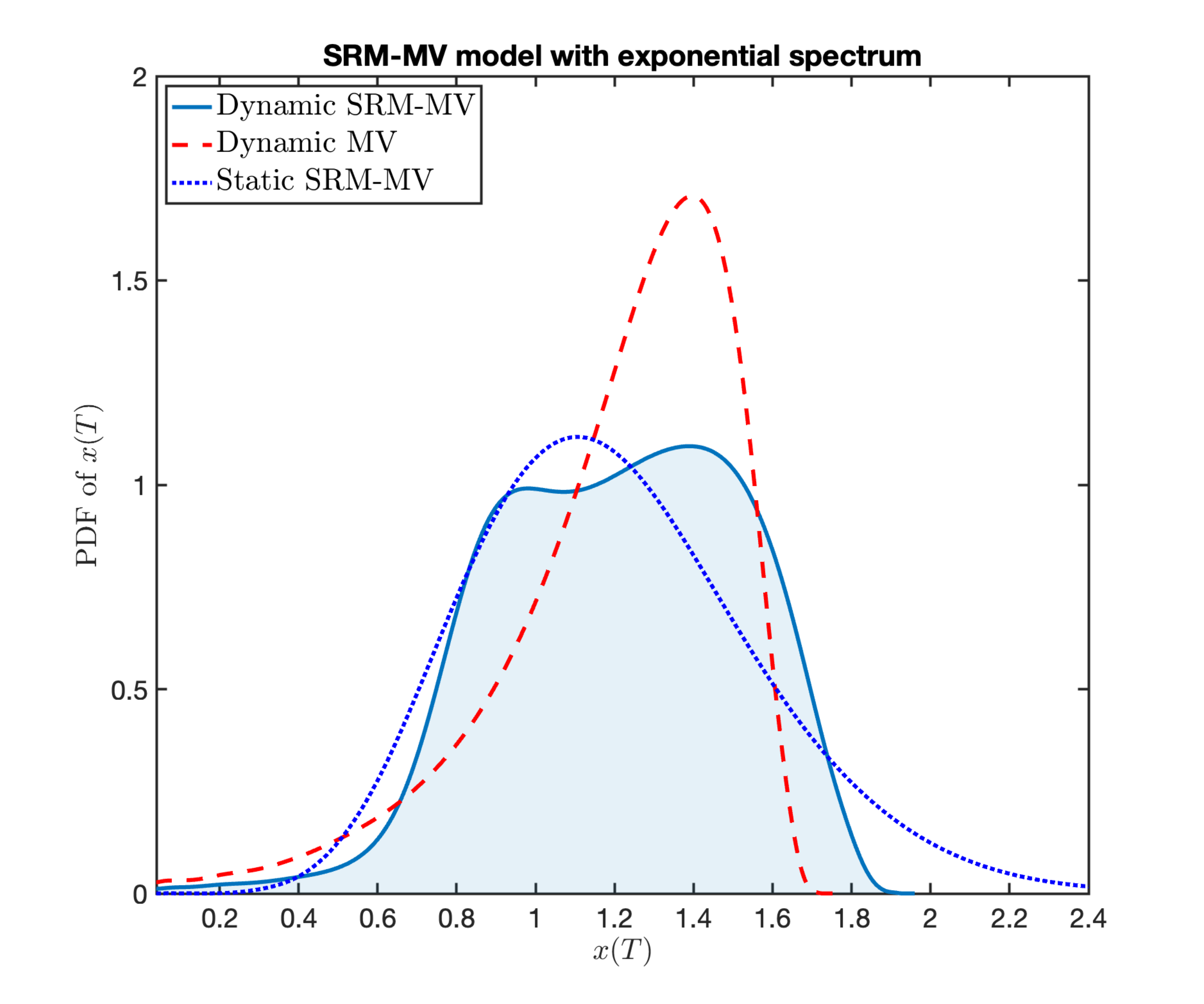}
   \caption{PDF of $x^*(T)$: ($\cP_{\srmv}$) with exponential spectrum }
   \label{fig:pdf_srmv_exp}
\end{subfigure}
\begin{subfigure}{0.49\textwidth}
   \includegraphics[width=\textwidth]{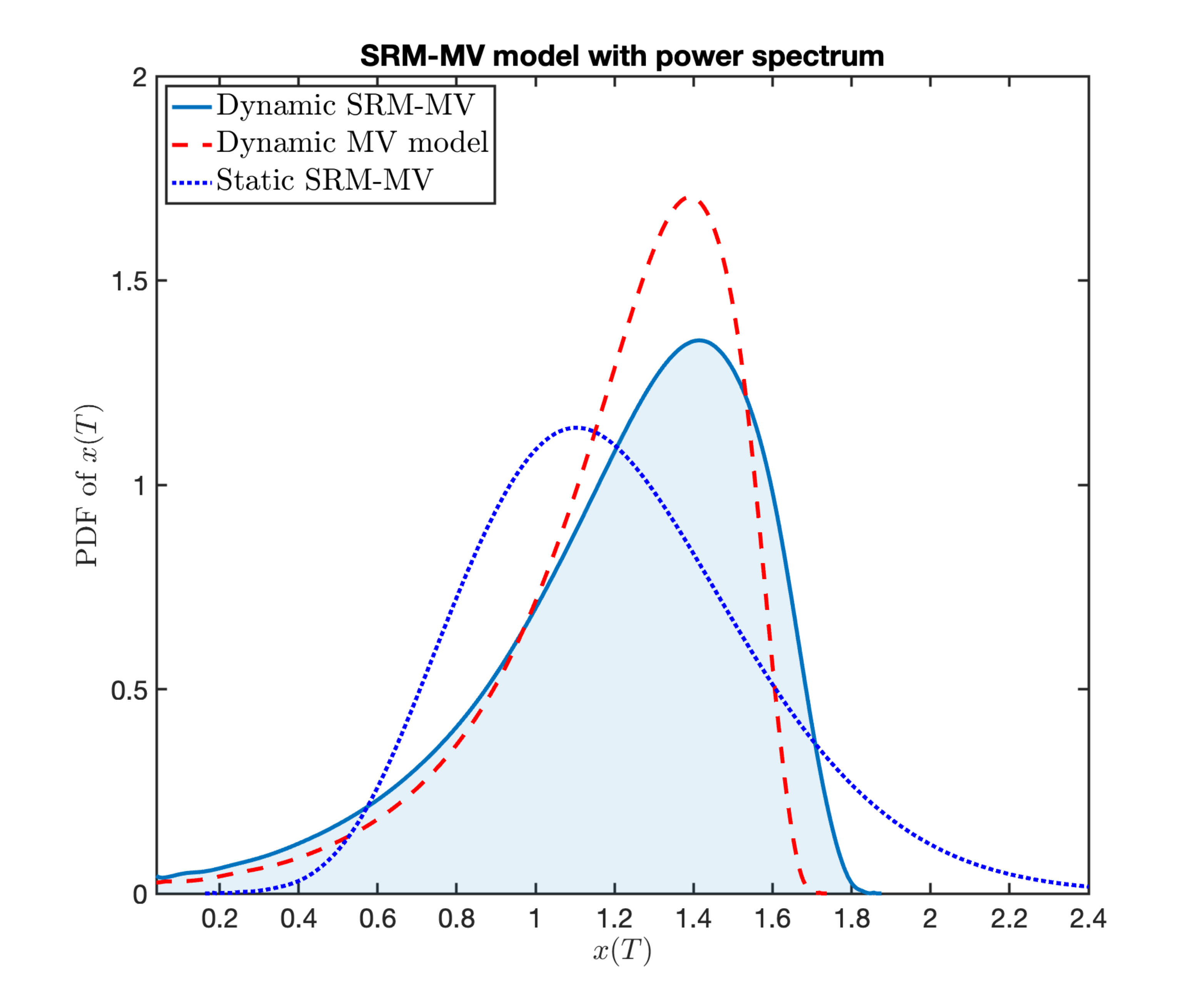}
   \caption{PDF of $x^*(T)$: ($\cP_{\srmv}$) with power spectrum }
   \label{fig:pdf_srmv_pow}
\end{subfigure}
\begin{subfigure}{0.49\textwidth}
   \includegraphics[width=\textwidth]{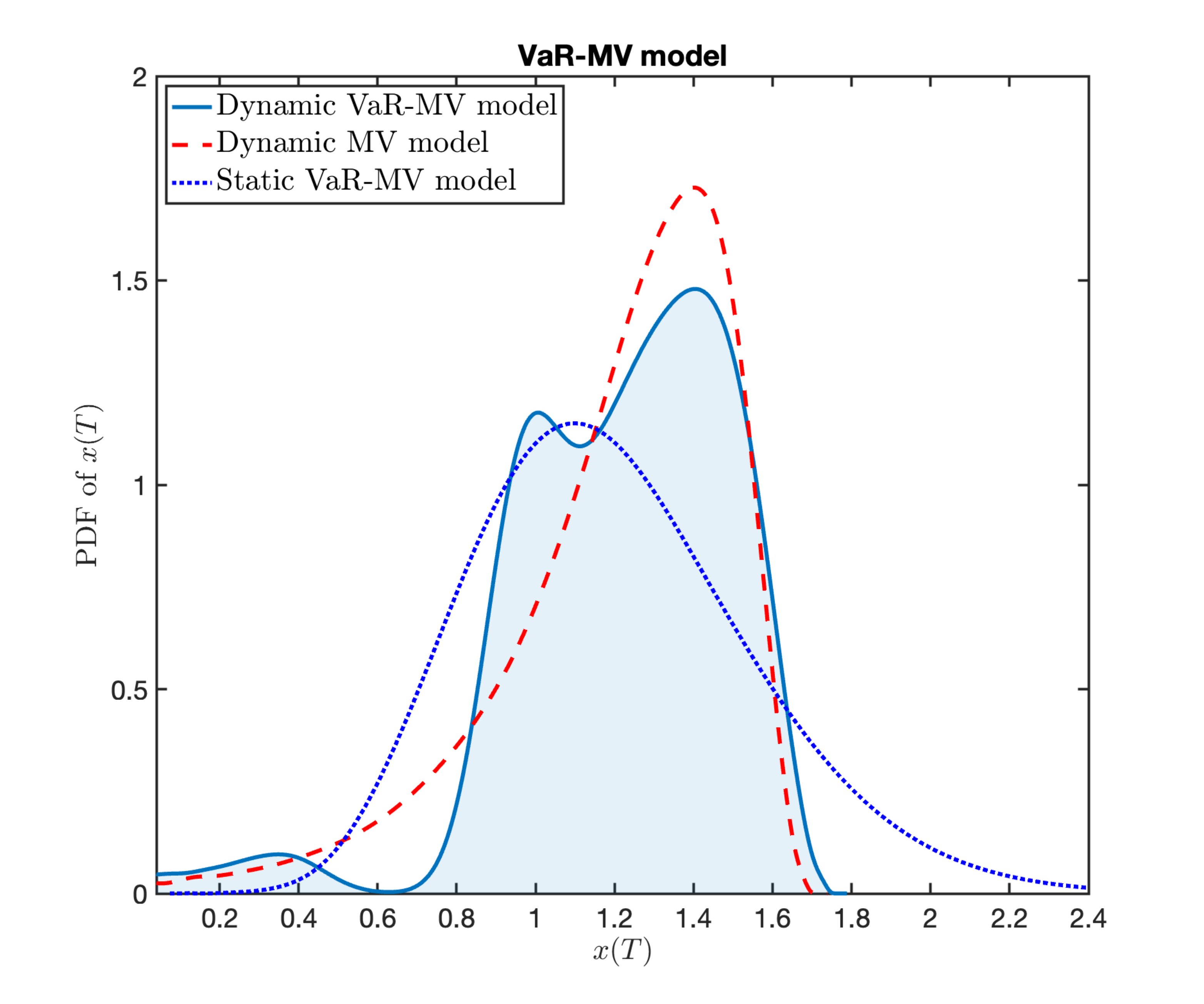}
   \caption{PDF of $x^*(T)$: ($\cP_{\vrmv}$)}
   \label{fig:pdf_vrmv}
\end{subfigure}
\caption{Empirical PDF of terminal wealth generated by different models} \label{fig:PDF_T=1}
\end{figure}

Once we have the samples of terminal wealth for each of these policies, we plot the empirical probability density (PDF) of $x^*(T)$. Figure \ref{fig:PDF_T=1} compares the PDF of various policies with parameters set to $\omega_{\srmv}=\omega_{\vrmv}=0.3$, $k_e=10$, $k_p=0.6$, $\gamma=10\%$, and initial wealth and target wealth of $x_0=1$ and $x_d=1.2$, respectively. In all figures, we use the PDF of wealth generated by the dynamic MV policy as the benchmark (indicated by the red dashed line). In Figure \ref{fig:pdf_srmv_exp}, the shaded area represents the PDF generated by the model ($\cP_{\srmv}|{\exp}$), which deviates significantly from the PDF of the dynamic MV model. This deviation is desirable for investment. In the domain of gain, the model ($\cP_{\srmv}|_{\exp}$) has a higher probability of achieving a better gain, while in the domain of loss, it lowers the probability of loss compared to the MV model (i.e., see the area $x(T)<0.7$). The static SRM-MV policy produces a Gaussian-type distribution with a significantly larger variance than the dynamic SRM-MV model. However, in Figure \ref{fig:pdf_srmv_pow}, the power spectrum-based model ($\cP_{\srmv}|_{\pow}$) shows a different pattern than the exponential spectrum-based model. It only increases the probability in the domain of gain but does not lower the probability of loss. \footnote{The pattern of the power-spectrum based model is robust to changes in the model parameters. In an experiment that was not reported, we tested different parameter values and found that the basic pattern remained consistent.} Figure \ref{fig:pdf_vrmv} indicates that the VaR-MV model reduces the probability of loss as well, but it performs similarly to the dynamic MV model in the domain of gain.

The basic pattern illustrated in Figure \ref{fig:PDF_T=1} can be more accurately quantified using several performance measures. We present the variance, semivariance, Sharpe Ratio, Sortino Ratio (\cite{Sortino:2001}), $10\%$-VaR, $5\%$-VaR, and the Rachev ratio (\citealp{Biglova:JPM2004}) for terminal wealth being $x_d=1.2$ and $x_d=1.3$ in Tables \ref{Table_performance_r20} and \ref{Table_performance_r30}, respectively. For each performance measure (column), we use a color scale to indicate the rank, with green representing the best performance, red representing the worst, and yellow representing the middle-level. Of these measures, we place particular emphasis on the Sortino Ratio and Rachev Ratio. The Sortino Ratio measures risk-adjusted return using downside standard deviation, and is recognized as an improved version of the Sharpe Ratio for portfolio performance. The Rachev Ratio measures the ratio between the mean of the best $\alpha\%$ values and the worst $\alpha\%$ values of the terminal wealth, with a higher value being preferred. Tables \ref{Table_performance_r20} and \ref{Table_performance_r30} show that the dynamic SRM-MV model $(\cP_{\srmv})|_{\exp}$ provides the best Sortino Ratio, the best $10\%$-Rachev Ratio, and the lowest semivariance. It also offers relatively good performance on other measures. Note that, even when using similar objective functions, the dynamic SRM-MV model $(\cP_{\srmv})|_{\exp}$ performs significantly better than its static counterpart model $\cP_{\srmv}^{\static}|_{\exp}$. Additionally, if we only consider downside risk measures, the dynamic model $(\cP{\vrmv})$ has the best performance with respect to $5\%$-VaR and $10\%$-VaR compared to the other models. Tables \ref{Table_performance_r20} and \ref{Table_performance_r30} also demonstrate that the SRM-MV model $(\cP_{\srmv})|_{\pow}$ performs poorly. This observation indicates that designing a suitable spectrum in the SRM-MV model is crucial to achieving good performance.

\begin{table}[!h]
\centering
\resizebox{0.9\textwidth}{!}{%
\begin{tabular}{@{}lcccccccc@{}}
\toprule
Row & \multicolumn{1}{l}{Variance} & \multicolumn{1}{l}{Semi Vari} & \multicolumn{1}{l}{Sharpe} & \multicolumn{1}{l}{\textbf{Sortino} } & \multicolumn{1}{l}{$\VaR_{5\%}$} & \multicolumn{1}{l}{$\VaR_{10\%}$} & \multicolumn{1}{l}{$ \Rach_{5\%} $ } & \multicolumn{1}{l}{ $\Rach_{10\%}$ } \\
\toprule
$\cP_{\mv}$ & \cellcolor[HTML]{548235}0.091 & \cellcolor[HTML]{FFBD6B}0.073 & \cellcolor[HTML]{548235}0.564 & \cellcolor[HTML]{FF2621}0.628 & \cellcolor[HTML]{FFEB84}-0.601 & \cellcolor[HTML]{EBDF7B}-0.801 & \cellcolor[HTML]{FF2721}0.810 & \cellcolor[HTML]{FF1A1B}1.051 \\
\midrule
$\cP_{\srmv}^{\static}|_{\exp}$ & \cellcolor[HTML]{FF6E3E}0.200 & \cellcolor[HTML]{B3BC61}0.053 & \cellcolor[HTML]{FF1A1B}0.482 & \cellcolor[HTML]{FFEB84}0.938 & \cellcolor[HTML]{FFB264}-0.577 & \cellcolor[HTML]{FF5E35}-0.703 & \cellcolor[HTML]{AAB75D}2.161 & \cellcolor[HTML]{FFEB84}2.226 \\
$\cP_{\srmv}|_{\exp}$ & \cellcolor[HTML]{FFEB84}0.117 & \cellcolor[HTML]{548235}0.028 & \cellcolor[HTML]{FFEB84}0.499 & \cellcolor[HTML]{548235}1.015 & \cellcolor[HTML]{90A750}-0.826 & \cellcolor[HTML]{C9CA6B}-0.835 & \cellcolor[HTML]{FFEB84}2.064 & \cellcolor[HTML]{548235}2.528 \\
\midrule
$\cP_{\srmv}^{\static}|_{\pow} $ & \cellcolor[HTML]{FF0000}0.273 & \cellcolor[HTML]{FFEB84}0.072 & \cellcolor[HTML]{FF4D34}0.486 & \cellcolor[HTML]{EEE17C}0.946 & \cellcolor[HTML]{FF0000}-0.503 & \cellcolor[HTML]{FF0000}-0.650 & \cellcolor[HTML]{A3B25A}2.170 & \cellcolor[HTML]{F9E882}2.237 \\
$\cP_{\srmv}|_{\pow} $ & \cellcolor[HTML]{F3E47E}0.115 & \cellcolor[HTML]{FF4627}0.077 & \cellcolor[HTML]{B2BC61}0.529 & \cellcolor[HTML]{FF3327}0.648 & \cellcolor[HTML]{FF4F2C}-0.536 & \cellcolor[HTML]{FF9D58}-0.738 & \cellcolor[HTML]{FF3729}0.913 & \cellcolor[HTML]{FF2621}1.119 \\
\midrule
$\cP_{\vrmv}^{\static}$ & \cellcolor[HTML]{FFE480}0.122 & \cellcolor[HTML]{5F893A}0.031 & \cellcolor[HTML]{FFC872}0.496 & \cellcolor[HTML]{A2B259}0.980 & \cellcolor[HTML]{D5D170}-0.687 & \cellcolor[HTML]{FFEB84}-0.782 & \cellcolor[HTML]{548235}2.258 & \cellcolor[HTML]{C8CA6B}2.323 \\
$\cP_{\vrmv}$ & \cellcolor[HTML]{769745}0.096 & \cellcolor[HTML]{FF0000}0.078 & \cellcolor[HTML]{7D9B48}0.549 & \cellcolor[HTML]{FF1A1B}0.609 & \cellcolor[HTML]{548235}-0.950 & \cellcolor[HTML]{548235}-0.950 & \cellcolor[HTML]{FF1A1B}0.725 & \cellcolor[HTML]{FF3B2C}1.242\\
\bottomrule
\end{tabular}%
}
\caption{The performance of the different portfolio policies with a target expected return $r=20\%$ ($x_d=1.2$) }\label{Table_performance_r20}
\end{table}

\begin{table}[!h]
\centering
\resizebox{0.9\textwidth}{!}{%
\begin{tabular}{@{}lcccccccc@{}}
\toprule
Row & \multicolumn{1}{l}{Variance} & \multicolumn{1}{l}{Semi Vari} & \multicolumn{1}{l}{Sharpe} & \multicolumn{1}{l}{ \textbf{Sortino} } & \multicolumn{1}{l}{$\VaR_{5\%}$} & \multicolumn{1}{l}{$\VaR_{10\%}$} & \multicolumn{1}{l}{$ \Rach_{5\%} $ } & \multicolumn{1}{l}{ $\Rach_{10\%}$ } \\
\toprule
$\cP_{\mv}$ & \cellcolor[HTML]{548235}0.242 & \cellcolor[HTML]{FF0000}0.153 & \cellcolor[HTML]{548235}{0.544} & \cellcolor[HTML]{FF1A1B}0.683 & \cellcolor[HTML]{FF6639}-0.217 & \cellcolor[HTML]{FF0000}-0.579 & \cellcolor[HTML]{FF1A1B}0.953 & \cellcolor[HTML]{FF1A1B}1.116 \\
\midrule
$\cP_{\srmv}^{\static}|_{\exp}$ & \cellcolor[HTML]{FF9051}0.287 & \cellcolor[HTML]{FBE882}0.074 & \cellcolor[HTML]{FFE17F}0.503 & \cellcolor[HTML]{F2E37E}0.990 & \cellcolor[HTML]{FCE982}-0.506 & \cellcolor[HTML]{FFEB84}-0.652 & \cellcolor[HTML]{FFEB84}2.060 & \cellcolor[HTML]{FFEB84}2.130 \\
$\cP_{\srmv}|_{\exp}$ & \cellcolor[HTML]{FFEB84}0.282 & \cellcolor[HTML]{548235}0.069 & \cellcolor[HTML]{FDEA83}0.504 & \cellcolor[HTML]{548235}1.016 & \cellcolor[HTML]{548235}-0.658 & \cellcolor[HTML]{E0D875}-0.700 & \cellcolor[HTML]{CECD6E}2.119 & \cellcolor[HTML]{94AA53}2.256 \\
\midrule
$\cP_{\srmv}^{\static}|_{\pow} $ & \cellcolor[HTML]{FF9252}0.287 & \cellcolor[HTML]{FFEB84}0.074 & \cellcolor[HTML]{FFEB84}0.503 & \cellcolor[HTML]{F3E47F}0.990 & \cellcolor[HTML]{FFEB84}-0.504 & \cellcolor[HTML]{FFEB84}-0.652 & \cellcolor[HTML]{5F893A}2.254 & \cellcolor[HTML]{598538}2.326 \\
$\cP_{\srmv}|_{\pow} $ & \cellcolor[HTML]{8BA34E}0.255 & \cellcolor[HTML]{FF3C22}0.133 & \cellcolor[HTML]{E0D876}0.511 & \cellcolor[HTML]{FF3C2C}0.733 & \cellcolor[HTML]{FF8D4F}-0.301 & \cellcolor[HTML]{FF0805}-0.582 & \cellcolor[HTML]{FF3227}1.084 & \cellcolor[HTML]{FF3629}1.256 \\
\midrule
$\cP_{\vrmv}^{\static}$ & \cellcolor[HTML]{C6C869}0.269 & \cellcolor[HTML]{548235}0.069 & \cellcolor[HTML]{FFB96A}0.501 & \cellcolor[HTML]{FFEB84}0.988 & \cellcolor[HTML]{E9DE7A}-0.523 & \cellcolor[HTML]{F7E680}-0.665 & \cellcolor[HTML]{548235}2.266 & \cellcolor[HTML]{548235}2.332 \\
$\cP_{\vrmv}$ & \cellcolor[HTML]{FF0000}0.295 & \cellcolor[HTML]{FFB968}0.091 & \cellcolor[HTML]{FF1A1B}0.494 & \cellcolor[HTML]{FFA762}0.889 & \cellcolor[HTML]{FF0000}0.000 & \cellcolor[HTML]{548235}-0.920 & \cellcolor[HTML]{FF5639}1.275 & \cellcolor[HTML]{FFE782}2.110 \\ \bottomrule
\end{tabular}%
}
\caption{The performance of the different portfolio policies with a target expected return $r=30\%$ ($x_d=1.3$)}\label{Table_performance_r30}
\end{table}

\begin{figure}[!h]
\centering
\begin{subfigure}{0.44\textwidth}
   \includegraphics[width=\textwidth]{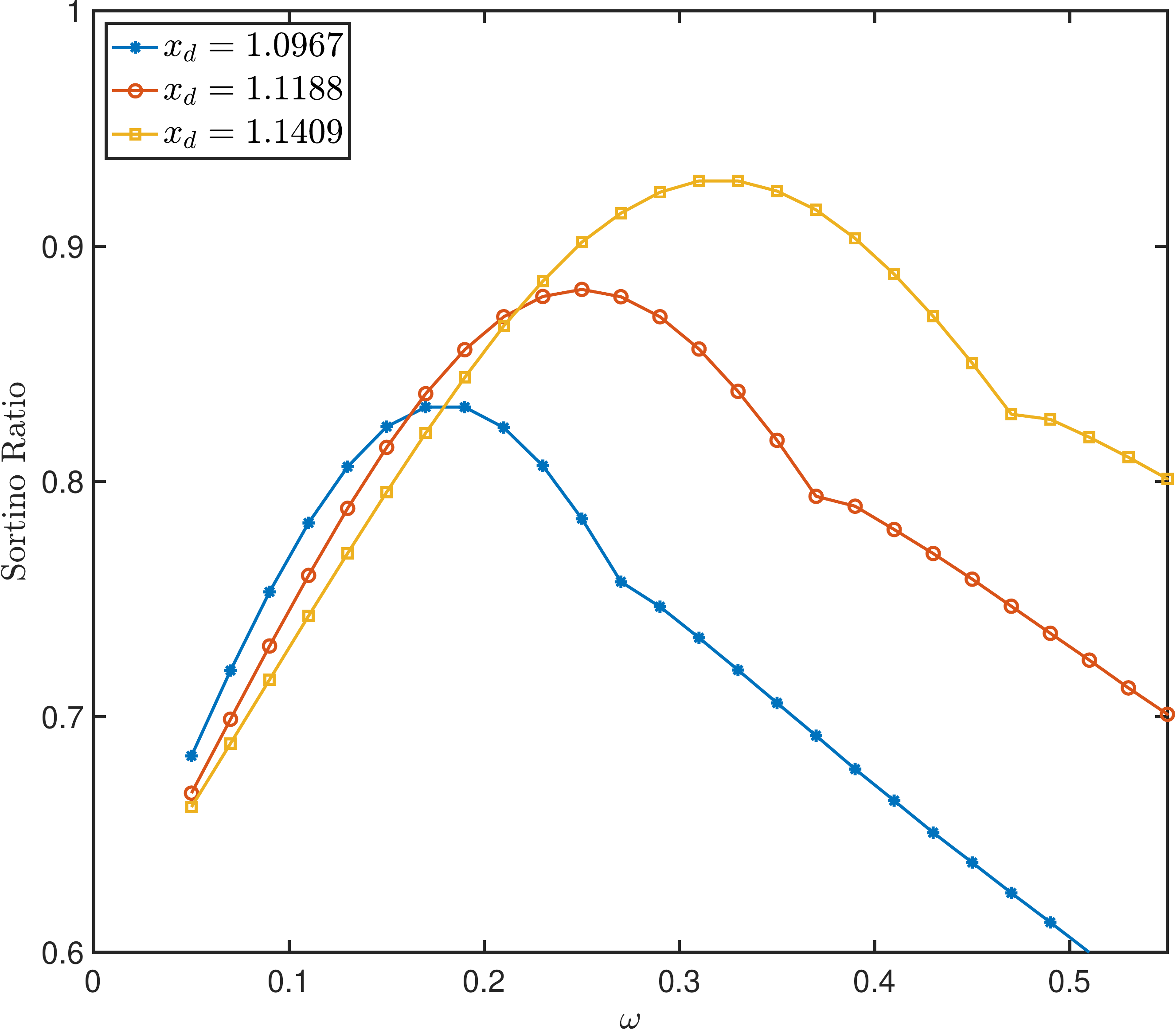}
   \caption{Sortino Ratio in model $(\cP_{\srmv})$}
   \label{fig:omega_sortino_srmv_exp}
\end{subfigure}
\begin{subfigure}{0.44\textwidth}
   \includegraphics[width=\textwidth]{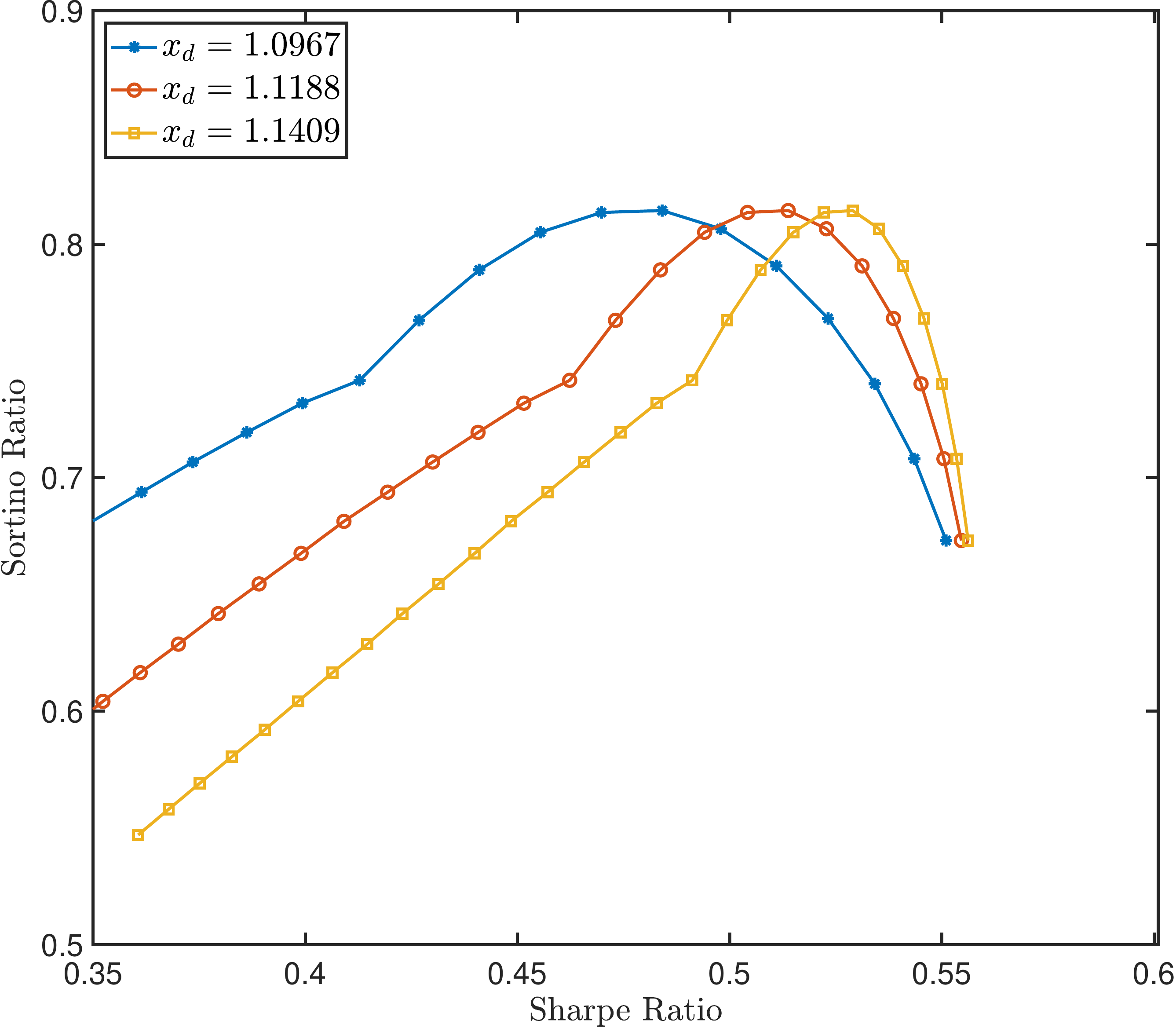}
   \caption{Sharpe Ratio and Sortino Ratio in model $(\cP_{\srmv})$ }
   \label{fig:sharpe_sortino_srmv_exp}
\end{subfigure}
\caption{Impact of weighting parameter in model ($\cP_{\srmv}$)} \label{fig:srmv_omega}
\end{figure}

We found that the dynamic model $(\cP_{\srmv})|_{\exp}$ performed the best in our tests. However, it is crucial to note that the choice of the weighting parameter $\omega_{\srmv}$ significantly affects the results. Figure \ref{fig:omega_sortino_srmv_exp} shows the Sortino ratio plotted against different values of $\omega_{\srmv}$ generated by the $(\cP_{\srmv}|_{\exp})$ model. The plot indicates that the Sortino ratio has a unimodal relationship with $\omega_{\srmv}$, suggesting an optimal value that maximizes the Sortino ratio. Furthermore, Figure \ref{fig:sharpe_sortino_srmv_exp} displays both the Sharpe Ratio and Sortino Ratio for various values of $\omega_{\srmv}$. It is important to note that these two measures are not consistent with each other, and the Sortino ratio decreases when the Sharpe ratio surpasses a specific threshold. Hence, it is crucial to carefully consider the choice of $\omega_{\srmv}$ to obtain reliable results.

\section{Conclusion} \label{sec_conclusion}

This paper investigates the continuous-time dynamic hybrid mean-variance (MV) portfolio optimization with spectral risk measure (SRM) and VaR. By utilizing the martingale approach with quantile formulation, we have successfully developed a solution to these problems. In contrast to the traditional dynamic MV model, the optimal portfolio policy generated by our hybrid model exhibits a distinct pattern. Specifically, it tends to hold more risky assets in both favorable and adverse market conditions, while holding less in the intermediate condition compared to the dynamic MV policy. This pattern leads to a desired distribution of terminal wealth. Our numerical test shows that the SRM-MV model with the exponential spectrum outperforms the benchmark model in terms of the Sortino ratio and downside risk measures. However, our current results are still insufficient in several aspects. First, these models need to be evaluated in real-world applications. To achieve this, the continuous-time policy must be translated into an implementable policy in a discrete-time setting. Second, since the variance and SRM or VaR are not separable in terms of dynamic programming, the current approach develops a pre-committed policy that is not time-consistent. Developing a time-consistent policy for these hybrid portfolio optimization models is a challenging future research task.


\appendix
\section{Proofs of Main Results }

\subsection{Proof of Theorem \ref{thm_srmv_xT}}\label{appendix_proof_thm_srmv_xT}
To solve problem ($\mathcal{G}_{\srmv}$), we introduce the Lagrange multipliers $\rho \in\bR $ and $ \eta\in\bR $ for the constraints (\ref{const_Gt_x0}) and (\ref{const_Gt_d}), respectively, which yields the following Lagrangian relaxation problem,
\begin{align*}
\hat{\cG}_{\srmv}(\rho,\eta):&~\min_{ G(\cdot)\in\bG }~\int_{0}^{1} G^{2}(s) ds - \omega_{\srmv} \int_{0}^{1} \psi(s)G(s) ds \\
& - \rho \int_{0}^{1} G(s) ds + \eta \int_{0}^{1} G(s) K_{0}^{-1}(1-s) ds.
\end{align*}
If there is no additional constraint except $G(\cdot)\in \bG$, checkin the first-order derivative of problem ($\hat{\cG}_{\srmv}(\rho,\eta)$) yields the optimal solution, $G^{\dag}(s)=\frac{\rho - \eta K_0^{-1}(1-s)+\omega_{\srmv}\psi(s)}{2}$ for all $s\in[0,1]$. To insure the constraint, $G(s) \geq 0$, we may compare $G^{\dag}(s)$ with $0$. Suppose $s^{\dag}$ is solution of the equation $\rho-\eta K_0^{-1}(1-s^{\dag}) + \omega_{\srmv}\psi(s^{\dag}) = 0$, which implies $G^{\dag}(s^{\dag})=0$. Under Assumption \ref{asmp_zt}, the function $G^{\dag}(s)$ is a non-decreasing function with respect to $s$. Thus it has $G^{\dag}(s)<0$ for any $s\in[0,s^{\dag})$. Then we can conclude that the optimal solution of problem ($\hat{\cG}_{\srmv}(\rho,\eta)$) is $G^*(s)=0$ for any $s \in[0,s^{\dag})$ and $G^*(s) = G^{\dag}(s)$ for any $s\in[s^{\dag},1]$.

We then show that the Eq. (\ref{equ_xd_sr}) admits no solution when $\eta^*\leq0$. Since $\rho^*$ $=$ $\eta^* z^{\dag}-\omega_{\srmv}\psi\big(1- K_0(z^{\dag})\big)$, Eq. (\ref{equ_xd_sr}) can be written as,
\begin{align*}
x_d =\E \left[ \left(\frac{\eta^* \big( z^{\dag}-z(T)\big) - \omega_{\srmv} \Big( \psi(1- K_0(z^{\dag}))-\psi(1 - K_0(z(T)))\Big)}{2} \right)\1_{  \left\{ 0 < z(T) \leq z^{\dag} \right\} } \right].
\end{align*}
From the fact that $K_0(\cdot)$ is a non-decreasing function and $\psi(\cdot)$ is a non-increasing function, the right-hand side of the above equation is always non-positive since $z(T)\leq z^{\dag}$ and $\eta^*\leq0$. Thus, Eq. (\ref{equ_xd_sr}) has no solution when $\eta^*\leq0$ which further implies $\eta^*>0$. Once we have the optimal quantile function $G^*(\cdot)$ for the problem $(\cG_{\srmv})$, we can identify the optimal wealth $x_{\srmv}^*(T)$ by using Theorem B1 in \cite{JinZhou:2008}, i.e., it has $x_{\srmv}^*(T)=G^*(1-K_0(z(T)))$ which leads to the result (\ref{def_srmv_xT}).
\qedsymbol{}

\subsection{Proof of Theorem \ref{thm_vrmv_xT}}\label{appendix_proof_vrmv_xT}
We first solve the problem ($\cG_{\vrmv}$). We introduce the quantile function,
\begin{align}
G^*(s; \eta, \rho, \beta)=\frac{\rho-\eta K_0^{-1}(1-s)}{2} \1_{ \{L_1 <s\leq \gamma\}}
-\beta\1_{\{\gamma< s\leq L_1 \}} + \frac{\rho-\eta K_0^{-1}(1-s)}{2} \1_{ \{L_2 <s\leq 1\}}\label{def_G_opt_all}
\end{align}
for $\eta,\rho \in \mathbb{R}$ and $\beta\in[\underline{\beta},0]$, where $L_1$ and $L_2$ are defined as,
\begin{align}
L_1\triangleq\min\{1 - K_0\left(\frac{\rho}{\eta} \right), \gamma\},~~L_2\triangleq\max\Big\{ 1-K_0\left(\frac{\rho+2\beta}{\eta} \right), \gamma\Big\}.
\label{def_L1L2}
\end{align}

Then we show that $G^*(s;\eta^*,\rho^*,\beta^*)$ is the optimal solution of problem $(\cG_{\vrmv})$, if the following two equations,
\begin{align}
&\int_{L_1}^{\gamma} (\rho^*-\eta^* K_0^{-1}(1-s))\cdot K_0^{-1}(1-s)ds
  -2\beta \int_{\gamma}^{L_2}K_0^{-1}(1-s)ds\notag \\
&~~~~~~~~~~~~~~~~~~~~~~~~~~~~~~~~~~+ \int_{L_2}^{1} (\rho^*-\eta^* K_0^{-1}(1-s) ) K_0^{-1}(1-s)ds = 2 x_0,  \label{equ1}\\
& \int_{L_1}^{\gamma}(\rho^* - \eta^* K_0^{-1}(1-s)) ds
-2\beta^*(L_2-\gamma)+ \int_{L_2}^{1}(\rho^*-\eta^* K_0^{-1}(1-s))ds =2 x_d,\label{equ2}
\end{align}
admit the solution $\rho^*$ and $\eta^*>0$, and $\beta^*$ is the minimizer of the following problem,
\begin{align}
\beta^*=&\arg\min_{\beta\in[\underline{\beta},0]}~\left\{\int_0^1 G^*(s;\eta^*,\rho^*,\beta)^2 ds + \omega_{\vrmv} \beta \right\}. \label{opt_G}
\end{align}

Since the objective function (\ref{G_mvv_objective}) is the summation of a functional term $\int_0^1 G^2(s)ds$ and a function value $G(\gamma)$, we may solve this problem by a two-step scheme, i.e., we first solve the problem by setting $G(\gamma)=-\beta$ for some fixed value $\beta \in[\underline{\beta},0]$ and then we identify the optimal $\beta$ which minimizes the objective function.\footnote{Note that, $\beta$ has the feasible range as $\beta\in(\underline{\beta}, 0)$.} For given $\beta$, the problem $(\cG_{\vrmv})$ becomes
\begin{align*}
\hat{\cG}(\beta):~&~\ \min_{G(\cdot) \in \bG}~ \int_{0}^{1} G^{2}(s) ds - \omega_{\vrmv} \cdot G(\gamma) \notag \\
(s.t.)~~&~G(\cdot)~\textrm{satisfies (\ref{const_Gt_x0}),~(\ref{const_Gt_d}),~(\ref{const_Gt_bankruptcy})},\\
      ~~&~-G(\gamma) =\beta.
\end{align*}
Introducing the Lagrange multipliers $\eta \in \bR$ and $\rho \in \bR$ for the constraints (\ref{const_Gt_x0}) and (\ref{const_Gt_d}), respectively, yields the following partially relaxed problem (after ignoring the constant $\omega_{\vrmv} \beta$),
\begin{align*}
\mathcal{L}(\eta,\rho,  \beta):~\min_{G(\cdot)\in\bG}~&~\int_{0}^{1} G^{2}(s) ds - \rho \int_{0}^{1} G(s) ds + \eta \int_{0}^{1} G(s) K_{0}^{-1}(1-s) ds \\
(s.t.)~&~~G(\gamma) = -\beta, \\
      ~&~~G(s)\geq 0~~\textrm{for all}~~0 \leq s \leq 1.
\end{align*}
For the convenience of illustration, we first assume $\eta>0$.\footnote{The case of $\eta\leq0$ can never hold true, which can be verified by using the similar method of {the} problem ($\cP_{srmv}$).} Clearly, the optimal solution of problem $\hat{\cG}(\beta)$ is a feasible solution of problem $\cL(\eta,\rho,\beta)$. Thus, the weak duality property holds. On the other hand, if the optimal solution of the relaxed problem $\mathcal{L}(\eta,\rho,\beta)$ is also feasible to problem $\hat{\cG}(\beta)$, then such a solution is also the optimal solution of problem $\hat{\cG}(\beta)$. To employ such a duality relationship, we first solve problem $\mathcal{L}(\eta,\rho,\beta)$ by the decomposition approach.\footnote{As problem $\mathcal{L}(\eta,\rho,\beta)$ is a functional optimization problem, the typical solution method is the classical calculus of variation. However, due to its special structure, we may characterize the solution directly by decomposition.} Notice that $G(\cdot)$ is a right-continuous and nondecreasing function which implies that $G(s) < G(\gamma)$ for $s\in [0,\gamma)$ and $G(\gamma) \leq G(s)$ for $s \in [\gamma, 1] $. This property motivates us to decompose problem ($\mathcal{L}(\eta,\rho,\beta)$) into the following two subproblems,
\begin{align*}
\cL^1(\eta,\rho,\beta):~~\min_{G(\cdot)\in\bG}~&~ \int_{0}^{\gamma} G^{2}(s) ds - \int_{0}^{\gamma} G(s)
\Big(\rho-\eta K_{0}^{-1}(1-s)\Big) ds \\
(s.t.)~&~G(\gamma) = -\beta, \\
      ~&~G(s)\geq0,~\textrm{for all}~0\leq s<\gamma,
\end{align*}
and
\begin{align*}
\cL^2(\eta,\rho,\beta):~~\min_{G(\cdot)\in\bG}~&~ \int_{\gamma}^{1} G^{2}(s) ds - \int_{\gamma}^{1} G(s)\Big(\rho-\eta K_{0}^{-1}(1-s)\Big) ds \\
(s.t.)~&~G(\gamma) = -\beta, \\
      ~&~G(s)\geq -\beta,~\textrm{for all}~\gamma\leq s\leq1.
\end{align*}
Note that the integral kernel in the objective functions of these two subproblems can be written as
\begin{align*}
G^2(s)-G(s)\big(\rho - \eta K^{-1}_0(1-s) \big) &= \left( G(s) -\frac{\rho - \eta K^{-1}_0(1-s)}{2} \right)^2
                        -\frac{ \big(\rho - \eta K^{-1}_0(1-s)\big)^2 }{4}.
\end{align*}
The above completion of the square implies that, if there are no additional constraints except $G(\cdot)\in \mathbb{G}$, both of the two subproblems admit the optimal solution $G^{\ddag}(s,\rho,\eta)\triangleq\frac{\rho-\eta K_0^{-1}(1-s)}{2}$.
When the constraints exist, we may identify the optimal solution by comparing $G^{\ddag}(s,\rho,\eta)$ with the boundaries of $G(s)$, i.e., $0$ and $-\beta$. Under the Assumption \ref{asmp_zt}, the function $G^{\ddag}(s,\rho,\eta)$ is a non-decreasing function with respect to $s$, thus it has two threshold points,
\begin{align}
h_1: = 1-K_0\left( \frac{\rho}{\eta} \right)~\textrm{and}~~h_2\triangleq 1- K_0\left( \frac{\rho + 2\beta}{\eta} \right),\label{def_h1h2}
\end{align}
which satisfy {$G^{\ddag}(h_1,\rho,\eta)=0$ and $G^{\ddag}(h_2,\rho,\eta)=-\beta$. As $\beta<0$, it always has $0 \leq h_1<h_2 \leq 1$. Moreover, the solutions of the above subproblems also depend on the position of $\gamma$ in the interval $[0,1]$. Without loss of generality, we first assume $\gamma\in (h_1, h_2)$. The other cases, i.e., $\gamma \in [0, h_1]$ or $\gamma\in [h_2,1]$, will be examined at the end of the proof.

As for subproblem ($\cL^1(\eta,\rho,\beta)$), under this assumption $\gamma\in (h_1, h_2)$, it has $G^{\ddag}(s,\rho,\eta)<0$} for $s\in[0,h_1)$ and $G^{\ddag}(s,\rho,\eta) \geq 0$ for $s\in [h_1,\gamma)$. Thus, the optimal solution of problem ($\cL^1(\eta,\rho,\beta)$) is
\begin{align}
G^*(s,\rho,\eta)= G^{\ddag}(s, \rho, \eta)\1_{\{h_1\leq s<\gamma)\}}. \label{def_L1_G}
\end{align}
We then check subproblem ($\cL^2(\eta,\rho,\beta)$). Similarly, it has $G^{\ddag}(s,\rho,\eta)<-\beta$ if $s\in \big[\gamma,h_2\big)$ and $G^{\ddag}(s,\rho,\eta)\geq -\beta$ if $s\in \big[h_2,1 \big]$ which implies that, the function
\begin{align}
G^*(s,\rho,\eta)= -\beta\1_{\{ \gamma \leq s<h_2 \}}+ G^{\ddag}(s,\rho,\eta)\1_{ \{ h_2\leq s \leq 1 \} }\label{def_L2_G}
\end{align}
is the optimal solution of problem ($\cL^2(\eta,\rho,\beta)$). Combining the above two solutions (\ref{def_L1_G}) and (\ref{def_L2_G}) gives the solution of problem $\cL(\rho, \eta,\beta)$ as
\begin{align}
G^*(s,\rho, \eta)=G^{\ddag}(s, \rho, \eta)\1_{\{ h_1 \leq s < \gamma \} }
-\beta \1_{\{\gamma \leq s < h_2 \}}+
G^{\ddag}(s, \rho, \eta)\1_{\{h_2\leq s \leq1\} }.\label{def_G_opt_normal}
\end{align}
Due to the weak-duality, if there exist $\rho^*$ and $\eta^*$ such that the solution (\ref{def_G_opt_normal}) is feasible to constraints (\ref{const_Gt_x0}) and (\ref{const_Gt_d}), then the solution (\ref{def_G_opt_normal}) is the optimal solution of problem ($\hat{\cG}(\beta)$). Substituting (\ref{def_G_opt_normal}) to constraints  (\ref{const_Gt_x0}) and (\ref{const_Gt_d}) gives the equations for $\rho^*$ and $\eta^*$,
\begin{align*}
& \int_{h_1}^{\gamma} G^{\ddag}(s,\rho^*,\eta^*)K_0^{-1}(1-s)ds
  -\int_{\gamma}^{h_2} \beta K_0^{-1}(1-s)ds\\
&~~~~~~~~~~~~~~~~~~~~~~~~~~~~~~~~+ \int_{h_2}^{1} G^{\ddag}(s,\rho^*,\eta^*)K_0^{-1}(1-s)ds =x_0, \\
& \int_{h_1}^{\gamma} G^{\ddag}(s,\rho^*,\eta^*) ds
  -\int_{\gamma}^{h_2} \beta ds+ \int_{h_2}^{1} G^{\ddag}(s,\rho^*,\eta^*) ds = x_d.
\end{align*}
Clearly, the above two equations are just special cases of equations (\ref{equ1}) and (\ref{equ2}) when $\gamma \in(h_1,h_2)$. Recall that the above solution scheme solves the problem ($\hat{\cG}(\beta)$) for fixed $\beta\in [\underline{\beta},0]$. To identify the optimal $\beta$, we may vary $\beta$ and solve the problem $(\hat{\cG}(\beta))$, i.e., the optimal $\beta^*$ can be identified by
\begin{align}
\beta^*=\arg \min_{ \beta\in [\underline{\beta},0]} \int_0^1 ( G^*(s,\rho^*,\eta^*) )^2 ds + \omega_{\vrmv} \beta, \label{beta_opt}
\end{align}
where $G^*(s,\rho^*,\eta^*)$ is optimal solution of problem $(\hat{\cG}(\beta))$.

In the previous analysis, we have assumed that $\gamma \in (h_1, h_2)$. We then consider the other two cases: (i) $\gamma \in [0,h_1]$ and (ii) $\gamma \in [h_2,1]$.  For case (i), it has $G^{\ddag}(s,\rho,\eta)<0$ for $s\in [0,\gamma)$, which implies that the optimal solution of problem ($\cL^1(\eta,\rho,\beta)$) is $G^*(s,\rho,\eta)=0$ for $s \in[0,\gamma)$. As for the problem ($\cL^2(\eta,\rho,\beta)$), since $G^{\ddag}(s, \rho, \eta)<-\beta$ if $s\in [\gamma, h_2]$ and $G^{\ddag}(s,\rho, \eta)\geq -\beta$ if $s\in(h_2,1]$, the optimal solution is $G^*(s,\rho,\eta)=-\beta$ for  $s\in[\gamma, h_2)$ and {$G^*(s,\rho,\eta)=G^{\ddag}(s,\rho,\eta)$} if $s\in (h_2, 1]$. As a summary, if $\gamma \in [0,h_1]$, the optimal solution of problem $\cL(\eta,\rho,\beta)$ is
\begin{align}
{G^*(s,\rho,\eta)}=-\beta\1_{\{ \gamma \leq s< h_2\} } + G^{\ddag}(s,\rho,\eta)\1_{ \{h_2\leq  s\leq  1 \} }.\label{def_G_opt_deg1}
\end{align}
Clearly, the solution (\ref{def_G_opt_deg1}) is a degenerated case of (\ref{def_G_opt_all}). Indeed, since $\gamma<h_1$, the second piece of function in (\ref{def_G_opt_all}) does not exist. For the second degenerated case, $\gamma>h_2$, we may conduct {a similar} analysis. We omit the detail. The optimal solution of problem $\cL(\eta,\rho,\beta)$ is
\begin{align}
G^*(s,\rho,\eta) =G^{\ddag}(s,\rho,\eta)    \1_{\{h_1\leq  s\leq 1\} }.\label{def_G_opt_deg2}
\end{align}
The above solution is a special case of (\ref{def_G_opt_all}), i.e., when $\gamma>h_2$, the last three cases in (\ref{def_G_opt_all}) merge {into} one piece. In both of the above two special cases, after we solve {$G^*(s;\eta,\rho,\beta)$}, we may identify $\eta$ and $\rho$ by solving equations (\ref{equ1}) and (\ref{equ2}), and identify $\beta^*$ by similar method given in (\ref{beta_opt}). Once we have the optimal quantile $G^*(s;\eta^*,\rho^*,\beta^*)$ for problem $(\cG_{\vrmv})$, similar to problem $(\cA_{\srmv})$, we can translate the optimal quantile function to the {correspondent} optimal terminal wealth $x_{\vrmv}^*(T)$ by using Theorem B1 in \cite{JinZhou:2008}. This procedure leads to the result (\ref{def_vrmv_xT}).
\qedsymbol{}

\subsection{Detail in deriving solutions under Black-Scholes Market} \label{appendix_proof_srmv_vrmv_xt_BS}

Under the Assumption \ref{asmp_BS}, the SPD process $z(t)$ has the following result.
\begin{lemma} \label{lem_EZ_BS}
Considering $z(t)$ defined in (\ref{def_zt}), given $a, b \in \mathbb{R}$, $q_1, q_2 \in \mathbb{R}_+$ with $q_1\leq q_2$, it has
\begin{align}
&\E\Big[ \frac{z(T)}{z(t)}\big( a + b z(T)\big) \1_{  \{q_1 \leq z(T)\leq q_2\} } \Big|\cF_t \Big] \notag\\
&= a A(t)\big( \Phi\big( k_2(t) \big) - \Phi\big( k_1(t)  \big) \big)
+ b z(t) B(t) \big(\Phi( k_2(t) - v(t))-\Phi( k_1(t)- v(t)) \big), \label{lem_EZ_BS_equ}
\end{align}
where the parameters $m(t)$, $v(t)$, $A(t)$ and $B(t)$ are defined in Section \ref{sse_srmv_BS}, respectively; and
\begin{align}
k_1 &\triangleq  \frac{\ln( q_1/z(t))- m(t)}{v(t)}-v(t),~~k_2 \triangleq  \frac{\ln( q_2/z(t))- m(t)}{v(t)}-v(t).\label{def_k1k2}
\end{align}
\end{lemma}
The proof of Lemma \ref{lem_EZ_BS} is similar to the one in Proposition 7.1 in \cite{GaoXiongLi:2016}. Thus, we omit the detail.

We then derive the optimal wealth {process} for {the} problem $(\cP_{\srmv})$. Substituting (\ref{def_srmv_xT}) into (\ref{def_xt_expectation}) gives,
\begin{align*}
x_{\srmv}^*(t) &= \E\Big[ \frac{z(T)}{z(t)}x_{\srmv}^*(T) \big| \cF_t  \Big]\\
&=\frac{1}{2}\E\Big[\frac{z(T)}{z(t)}\big( \rho^* -\eta^* z(T))\big| \cF_t\Big]
+ \frac{\omega_{\srmv}}{2}\E\big[ e^{\ln \left( \frac{z(T)}{z(t)} \right)} \psi\big( 1-K_0(z(t)e^{\ln\left( \frac{z(T)}{z(t)} \right)}) \big) \big| \cF_t\big].
\end{align*}
In the above equation, we apply Lemma \ref{lem_EZ_BS} to the first term and write out the integration with respect to $\ln(z(T)/z(t))\sim \mathcal{N}(m(t),v^2(t))$ in the second term, which gives Eq. (\ref{def_srmv_xt_BS}). Similarly, using Lemma \ref{lem_EZ_BS}, the three equations (\ref{equ_zdag_sr}), (\ref{equ_xd_sr}) and (\ref{equ_x0_sr}) can be written as (\ref{equ_zdag_sr_BS}), (\ref{equ_xd_sr_BS}) and (\ref{equ_x0_sr_BS}), respectively.

For model $(\cP_{\vrmv})$, combining the terminal wealth (\ref{def_vrmv_xT}) in (\ref{def_xt_expectation}) gives
\begin{align}
x_{\vrmv}^*(t) & = \frac{1}{2}\E \big[ \frac{z(T)}{z(t)}\big( \rho^*-\eta^* z(T)\big)\1_{ \{ K_0^{-1}(1-\gamma) < z(T) \leq C_1 \} } \big|\cF_t \big] + \E\big[ \frac{z(T)}{z(t)}(-\beta^*)\1_{ \{C_2 < z(T) \leq K_0^{-1}(1-\gamma)\} } \big|\cF_t \big]\notag\\
&+ \frac{1}{2}\E\big[ \frac{z(T)}{z(t)}\big( \rho^*-\eta^* z(T) \big)\1_{ \{ z(T) \leq C_2 \} } \big|\cF_t \big]. \label{proof_xt_1}
\end{align}
By applying lemma \ref{lem_EZ_BS} to each term of expression (\ref{proof_xt_1}), we obtain the optimal wealth process $x_{\vrmv}^*(t)$ given in (\ref{vrmv_bs_xt}). Once we obtain the analytical expression of $x_{\srmv}^*(t)$ and $x_{\vrmv}^*(t)$, the correspondent optimal portfolio policies can be computed by Eq. (\ref{BS_ut}).

\qedsymbol{}

\subsection{The solution for the static hybrid portfolio optimization model} \label{appendix_static}

This section reports the solution schemes for the static counterpart problems of the hybrid portfolio optimization models ($\cP_{\srmv}$) and ($\cP_{\vrmv}$). In the static model, as the policy is kept unchange in horizon $[0,T]$, we only need to introduce the decision variables at time $t=0$, i.e., we use $u=(u_1,\ldots,u_n)^{\top} \in \mathbb{R}^n$ and $u_f \in \mathbb{R}$ to denote the wealth allocation on $n$ risky assets and {risk-free} asset, respectively. We then use
\begin{align*}
R \triangleq \Big( \frac{S_1(T)}{S_1(0)}, \frac{S_2(T)}{S_2(0)},\ldots, \frac{S_n(T)}{S_n(0)}  \Big)^{\top}
\end{align*}
to denote the random return vector of the risky assets and use $R_f \triangleq \frac{S_0(T)}{S_0(0)}$ to denote the risk-free {for the} time period $0$ to $T$, respectively. Then the terminal wealth $x^{\sst}(T)$ resulted from this static portfolio policy is
\begin{align}
x^{\sst}(T)|_{u,u_f}= {R}^{\top} u + R_f u_f. \label{static_x_T}
\end{align}
Clearly, the wealth $x^{\sst}(T)$ {is a random} variable depending on the portfolio decision $u$ and $u_f$. To construct the static SRM-MV and VaR-MV portfolio optimization models, it is more convenient to use the discrete scenario-based approach (see \cite{Acerbi:2002portfolio,BenatiRizzi:2007}). We assume there are totally $N$ discrete scenarios of the random return $R$ and use $R_{(i)}\in \mathbb{R}^n$ to denote the $i$-th scenario (realization) of the random return vector for $i=1,2,\ldots, N$. These discrete {scenarios} of the returns can be achieved either {by random sampling} from the empirical distribution or by using the historical data of returns directly. Given portfolio policy $u$ and $u_f$, the correspondent discrete scenarios of the terminal wealth is $x^{\sst}_{(i)}(T) = (R_{(i)})^{\top} u + R_f u_f$ for $i=1,\ldots, N$. Let $\E[R]$ and $Q =\E[(R-\E[R])(R-\E[R])^{\top}]$ be the expected value and covariance matrix of the random return $R$, respectively. The expected value and variance of the terminal wealth are
\begin{align*}
\E[x^{\sst}(T)]= \E[R]^{\top}u + R_f u_f,~~\textrm{and}~~~\cV[x^{\sst}(T)]= u^{\top} Q u,
\end{align*}
respectively.

We then focus on {reformulating} the SRM in a linear functional form. Similar to \cite{Acerbi:2002portfolio,BenatiRizzi:2007}, we discretize the interval $[0,1]$ by taking $N+1$ points evenly as $s_i=\frac{i-1}{N}$ for $i=1,\ldots,N+1$. For given spectrum function $\phi(s)$, we define
\begin{align*}
\psi_i := \int_{s_{i}}^{s_{i+1}} \psi(s)ds\approx \frac{1}{N} \psi(s_{i})
\end{align*}
for $i=1,\ldots,N$ as the discretization of $\phi(\cdot)$, which also satisfies $\sum_{i=1}^{N} \phi_i=1$. Following the similar method by \cite{Acerbi:2002portfolio}, for given $u$ and $u_f$, the discretized {SRM} can be expressed as
\begin{align}
\cM_{\psi}[x^{\sst}(T)]= -\sum_{i=1}^{N} \psi_i \cdot x^{\sst}_{(i:N)}(T), \label{def_M_dis}
\end{align}
where $x^{\sst}_{(i:N)}(T)$ denotes the $i$-th smallest element of $x^{\sst}_{(i)}(T)$ for $i=1,\ldots, N$. In formulation (\ref{def_M_dis}), we need to sort the realizations of $x^{\sst}_{(i)}(T)$ for $i=1,\ldots,N$ in an ascending order. However, as the terminal wealth is {a random} variable {affected} by the decision variable $u$ and $u_f$, the order of $\{x^{\sst}_{(i)}(T)\}|_{i=1}^N$ is also affected by the portfolio decision, which prevents us {from using} the formulation (\ref{def_M_dis}) directly. Fortunately, this difficulty can be conquered by formulating the sorting procedure as an optimization problem with auxiliary variables (see the detail in Proposition $3.1$ in \cite{Acerbi:2002portfolio}). Specifically, we introduce the auxiliary variables $\nu=(\nu_1, \nu_2,\ldots, \nu_N)$ and define the following function
\begin{align}
Y(\nu, u, u_f)\triangleq  \sum_{j=1}^{N-1} \Delta \psi_j \Big( j \cdot \nu_j
- \sum_{i=1}^{N}\big( \nu_j - x^{\sst}_{(i)}(T)\big)^{+} \Big)- \psi_N \sum_{i=1}^{N} x^{\sst}_{(i)}(T), \label{def_Y}
\end{align}
where $\Delta \psi_i \triangleq \psi_{i+1} -\psi_i$ for $i=1,\ldots, N-1$ and $\Delta \psi_N \triangleq -\psi_N$. Then the SRM (\ref{def_M_dis}) can be expressed as
\begin{align}
\cM_{\psi}[x^{\sst}(T)] = \min_{\nu}~Y(\nu, u, u_f).\label{def_M_dis2}
\end{align}
Note that, in (\ref{def_M_dis2}), we do not need to sort the realizations of $x^{\sst}_{(i)}(T)$. Using this formulation, the static {MV-SRM} portfolio optimization problem can be written as,
\begin{align}
(\cP_{\srmv}^{\static})~:~&~\min_{\nu, u,u_f}~u^{\top}Qu + \omega^{\sst}_{\srmv} \cdot Y(\nu,u,u_f) \notag\\
(s.t.)~&~\E[R]^{\top} u + R_fu_f = x_{d}, \label{st_const_return}\\
      ~&~\sum_{k=1}^{n} u_k + u_f = x_0, \label{st_const_wealth}\\
     ~&~(R_{(i)})^{\top}u + R_f u_f \geq 0,~~i=1,\ldots,N, \label{st_const_nobank}
\end{align}
where $x_d>x_0 e^{r_f T}$ is the target wealth level, {$\omega^{\sst}_{\srmv}\geq 0$} is the weighting parameter balancing the importance between the variance and SRM. The last constraint is due to the no bankruptcy constraint in ($10$)(see the main text). In problem ($\cP_{\srmv}^{\static}$), the objective function (i.e., $Y(\nu, u, u_f)$) involves the piece-wise linear term $(\nu_j - x^{\sst}_{(i)}(T)\big)^{+}$, which can be represented by auxiliary variables $y_{i,j}$ for $i,j=1,\ldots,N$ and linear constraints. Using this reformulation, problem $(\cP_{\srmv}^{\static})$ can be written as follows,
\begin{align}
(\cP_{\srmv}^{\static})~:~&~\min_{\nu, u,u_f,y_{i,j}}~~u^{\top}Qu + {\omega^{\sst}_{\srmv}} \cdot \Big ( \sum_{j=1}^{N-1} \Delta \psi_j \big( j \cdot \nu_j - \sum_{i=1}^{N} y_{i,j} \big) \notag\\
&~~~~~~~~~~~~~~~~~~~~~~~~~~- \psi_N \big(\sum_{i=1}^{N} R_{(i)}^{\top} u + R_f u_f\big)\Big) \notag\\
(s.t.) ~&~\nu_j- R_{(i)}^{\top} u - R_f u_f \leq y_{i,j},~~~~i,j=1,\ldots,N, \notag\\
~~&~y_{i,j}\geq 0,~~~~i,j=1,\ldots,N \notag\\
~&~\textrm{$u$ and $u_f$ satisfies (\ref{st_const_return}), (\ref{st_const_wealth}), (\ref{st_const_nobank})}.\notag
\end{align}
The above formulation is a convex quadratic programming (QP) problem which can be solved by {a convex} QP solver such as \cite{gurobi}.

For the static counterpart problem of the dynamic MV-VaR portfolio optimization ($\cP_{\vrmv}$), we utilize a similar method proposed by \cite{BenatiRizzi:2007,Cesarone:mv-var-2021}. It is important to note that, to compute the VaR of the portfolio, we still need to address the ordered statistics of the terminal wealth. However, the VaR's risk spectrum is a Dirac delta function, which does not satisfy the non-increasing property. Therefore, it does not admit a representation formula similar to SRM (i.e., Eq. (\ref{def_M_dis2})).To tackle this, we adopt a discrete-scenario-based setting similar to the approach taken in problem $(\cP_{\srmv}^{\static})$. This enables us to reformulate the static counterpart of the MV-VaR portfolio optimization problem as follows:

\begin{align*}
(\cP_{\vrmv}^{\static})~&~\min_{z_{\gamma}, u, u_f}~~u^{\top}Q u + {\omega^{\sst}_{\vrmv}} \cdot z_{\gamma} \\
(s.t.)~&~\kappa_{i} = R_{(i)}^{\top}u+ R_f u_f, ~~i=1,\ldots,N,\\
      ~&~\frac{1}{N}\sum_{i=1}^N (1-{z_i}) \geq {1-\gamma} \\
      ~&~z_{\gamma} \geq  -\big( \kappa_{i} + M (1- z_i ) \big),~~\forall~~i=1,\ldots,N,\\
      ~&~\textrm{$u$ and $u_f$ satisfies (\ref{st_const_return}), (\ref{st_const_wealth}), (\ref{st_const_nobank})}.\notag
\end{align*}
Formulation ($\cP_{\vrmv}^{\static}$) is a mixed-integer quadratic programming problem.

%
%
%
%
%
%
%
%
%
%
%
%
%
%
%

\bibliographystyle{elsarticle-harv}
\bibliography{MV-spectral_2022}


\end{document}